\documentclass[a4paper,aps,floats,intlimits,pre,notitlepage,intlimits,superscriptaddress,9pt,floatfix,twopages,twocolumn,8pt]{revtex4-1}
\usepackage{amsmath,dsfont,amsfonts}
\usepackage{mathtools,multirow,xfrac}
\usepackage[caption=false]{subfig}
\usepackage{ragged2e} 
\DeclareCaptionJustification{justified}{\justifying}
\usepackage[english]{babel}
\usepackage{flushend}
\usepackage{hyperref}
\hypersetup{pdfauthor={Adomaityte Toshniwal Sicuro Zdeborova},pdftitle={Planted hypermatching},%
            colorlinks, linktocpage=true, pdfstartpage=3, pdfstartview=FitV,%
    breaklinks=true, pdfpagemode=UseNone, pageanchor=true, pdfpagemode=UseOutlines,%
    plainpages=false, bookmarksnumbered, bookmarksopen=true, bookmarksopenlevel=1,%
    hypertexnames=true, pdfhighlight=/O,%
    urlcolor=blue, linkcolor=blue, citecolor=red,
        }

\usepackage{pgfplots}
\usepackage[caption=false]{subfig}

\usepackage{lmodern}
\usepackage[T1]{fontenc}
\usepackage[cal=boondoxo]{mathalfa}

\tikzset{
  common/.style={draw,name=#1,node contents={},inner sep=0,minimum size=2},
  disc/.style={circle,common=#1},
  square/.style={rectangle,common={#1}},
}

\usepackage{tikz}
\usetikzlibrary{patterns,decorations,arrows,shapes.geometric,arrows,mindmap,backgrounds,positioning,automata,positioning,fit,backgrounds,positioning,arrows,matrix,calc}

\newcommand{\comprimi}{\medmuskip=0mu
\thinmuskip=0mu
\thickmuskip=0mu}

\newcommand{\N}{{\mathds N}}

\newcommand{\R}{{\mathds R}}

\newcommand{\bfm}{{\boldsymbol m}}

\newcommand{\bw}{{\boldsymbol w}}

\newcommand{\sX}{{\mathsf X}}

\newcommand{\sH}{{\mathsf H}}
\newcommand{\sOmega}{{\mathsf \Omega}}

\newcommand{\sK}{{\mathsf K}}
\newcommand{\sU}{{\mathsf U}}
\newcommand{\sY}{{\mathsf Y}}
\newcommand{\sZ}{{\mathsf Z}}

\DeclareMathOperator{\dd}{d}
\DeclareMathOperator{\e}{e}

\allowdisplaybreaks
\begin{document}
\title{Planted matching problems on random hypergraphs}
\author{Urte Adomaityte}
\address{Department of Mathematics, King's College London, United Kingdom}
\author{Anshul Toshniwal}
\address{SPOC lab, EPFL, Lausanne, Switzerland}
\author{Gabriele Sicuro}
\address{Department of Mathematics, King's College London, United Kingdom}
\author{Lenka Zdeborov\'a}
\address{SPOC lab, EPFL, Lausanne, Switzerland}
\date{\today}
\begin{abstract}
We consider the problem of inferring a matching hidden in a weighted random $k$-hypergraph. We assume that the hyperedges' weights are random and distributed according to two different densities conditioning on the fact that they belong to the hidden matching, or not. We show that, for $k>2$ and in the large graph size limit, an algorithmic first order transition in the signal strength separates a regime in which a complete recovery of the hidden matching is feasible from a regime in which partial recovery is possible. This is in contrast to the $k=2$ case where the transition is known to be continuous. Finally, we consider the case of graphs presenting a mixture of edges and $3$-hyperedges, interpolating between the $k=2$ and the $k=3$ cases, and we study how the transition changes from continuous to first order by tuning the relative amount of edges and hyperedges.
\end{abstract}
\maketitle

\section{Introduction}
The study of inference problems has attracted a growing interest within the statistical physics community working on disordered systems \cite{nishimori2001,mezard2009,zdeborova2016}. Statistical physics techniques have been successfully applied to the study of a plethora of inference problems \cite{Decelle2011,error_corr_codes,zdeborova2016}, inspiring powerful algorithms for their solution \cite{mezard2009,Donoho2009,Bayati2011b} and unveiling sharp thresholds in the achievable performances with respect to the signal-to-noise ratio in the problem. Such thresholds delimit regions in which recovery of the signal is information-theoretically impossible, or easy, or hard (i.e., information theoretically possible but not achievable, or suboptimally achievable, by known polynomial-time algorithms) \cite{RicciTersenghi2019}.

The planted matching problem has recently been an object of a series of works that unveiled a non trivial phenomenology. The interest in it stems from a practical application, namely particle tracking \cite{CKKVZ_2010}: in the particle tracking problem, each particle appearing in a snapshot taken at time $t$ has to be assigned to the corresponding image in the frame taken at previous time $t-\Delta t$ via a maximum likelihood principle. This setting can be reformulated as an inference problem on a complete bipartite graph, in which the hidden truth corresponds to a perfect matching, and each feasible particle displacement is associated to an edge linking two nodes representing the old and new positions, weighted with the likelihood corresponding to the displacement itself. The maximum likelihood assignment can be found efficiently, e.g., using belief propagation \cite{MDM_2008,Bayati2011}. In a simplified, but analytically treatable, setting, a series of recent works \cite{MMX_2019,SSZ_2020,ding2021} revisited the problem considering a random graph of $N$ vertices containing a hidden perfect matching characterised by an edge weight distribution $\hat p$ different from the distribution $p$ of all other edge weights. By means of theoretical methods developed for the study of the random-link matching problem \cite{mezard1986,aldous2001}, it was shown that a phase transition takes place with respect to a certain measure of similarity between the distributions $p$ and $\hat p$ when the system size $N$ is large. A regime in which the hidden structure can be recovered up to $O(1)$ edges (complete recovery) is separated from a regime in which only a finite fraction of the edges can be correctly identified (partial recovery). Moreover, the transition is found to be continuous and, for a specific choice of $p$ and $\hat p$, proven to be of infinite order. Interestingly, it has been shown, at the level of rigour of theoretical physics, that the phenomenology extends to the so-called planted $k$-factor problem \cite{Bagaria2020,SZ_2021}, in which the hidden structure is a $k$-factor of the graph, that is a $k$-regular sub-graph including all the nodes.

In this work we will investigate the planted matching problem on hypergraphs. In hypergraphs edges may have more than two associated nodes. This natural extension of graphs is particularly interesting as many applications involve multiple classes to be matched at the same time (e.g., in the case in which a customer has to be matched to multiple types of products) \cite{Battiston2021}.
The minimum matching problem on weighted hypergraphs consists in finding a set of hyperedges such that every node belongs to one hyperedge in the set and the total weight of the hyperedges is minimized. The planted matching problem on hypergraphs can be motivated by particle tracking in $k$ consecutive snapshots where the probability that a particle moved on a given path is a non-separable function of its $k$ positions. This will be the case for most dynamical processes with some kind of inertia, e.g., a particle is more likely to keep its direction of movement rather than change direction randomly. In this application the hypergraph is a fully connected $k$-partite graph where each possible trajectory of a single particle corresponds to a hyperedge. The actual trajectory of that particle is in the planted set of hyperedges.

We will thus study a planted matching on hypergraphs and show that such apparently minimal generalisation bears remarkable differences with respect to the planted matching problem.
In the considered setting, the `signal' will consist of a perfect matching within a given graph, in which nodes are grouped in $k$-plets, each one bearing a weight distributed with density $\hat p$. Hyperedges not belonging to the hidden structure have weights distributed with density $p$. As in the planted matching problem, the goal is to recover the signal from the observation of the weighted hypergraph. 

The paper is organised as follows. We focus on a specific ensemble of hypergraphs, introduced in Section \ref{sec:ens}, where we specify the rules used to construct a random hypergraph with a hidden (or planted) matching within this ensemble. In Section \ref{sec:bp} we describe the belief propagation algorithm for the estimation of the marginals of the posterior probability: the algorithm relies on the knowledge of the construction rules given in Section \ref{sec:ens}. The performance of the algorithm is then investigated with respect to two estimators, namely the (block) maximum-a-posteriori matching and the so-called symbol maximum-a-posteriori estimator, i.e., the set of hyperedges whose marginal probability of belonging to the hidden matching is larger than $\sfrac{1}{2}$. In Section \ref{sec:transit} we show, by means of a probabilistic analysis of the belief propagation equations, that an algorithmic transition occurs between a phase with partial recovery of the signal and a phase with full recovery of the signal. The transition is found, for $k>2$, to be of first order, unlike the aforementioned $k=2$ case. A mixed model, involving both edges and hyperedges, is introduced in Section \ref{sec:mixed}: it is shown that the first order transition becomes of second order when a finite fraction of edges are introduced in the hypergraph. Finally, in Section \ref{sec:conclusions} we give our conclusions.

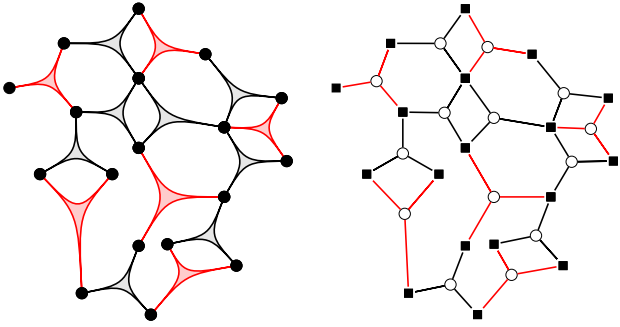
\begin{figure}
    \begin{center}
{ \newcommand{\hyperedgeb}[7][]
{
\draw[#1]($(#2)$) to [out=#5,in=#6] ($(#3)$) to [out=#6, in=#7] ($(#4)$) to [out=#7, in=#5] ($(#2)$);
\node [circle,draw,fill=black,inner sep=2] () at ($(#2)$) {};
\node [circle,draw,fill=black,inner sep=2] () at ($(#3)$) {};
\node [circle,draw,fill=black,inner sep=2] () at ($(#4)$) {};
}

\newcommand{\hyperedge}[4][]
{
\node (C) at (barycentric cs:#2=1,#3=1,#4=1) {};
\draw[#1]($(#2)$) .. controls ($(C)$) and ($(C)$) .. ($(#3)$) .. controls ($(C)$) and ($(C)$) .. ($(#4)$) .. controls ($(C)$) and ($(C)$) .. ($(#2)$);
\node [circle,draw,fill=black,inner sep=2] () at ($(#2)$) {};
\node [circle,draw,fill=black,inner sep=2] () at ($(#3)$) {};
\node [circle,draw,fill=black,inner sep=2] () at ($(#4)$) {};
}

\newcommand{\hyperedgeF}[4][]
{
\node[circle,draw,fill=white,inner sep=2] (C) at (barycentric cs:#2=1,#3=1,#4=1) {};

\draw[#1,fill=none](#2) -- (C) -- (#3) -- (C) -- (#4);
\node [rectangle,draw,fill=black,inner sep=2] () at ($(#2)$) {};
\node [rectangle,draw,fill=black,inner sep=2] () at ($(#3)$) {};
\node [rectangle,draw,fill=black,inner sep=2] () at ($(#4)$) {};
}

\tikzset{
    position/.style args={#1:#2 from #3}{
        at=(#3.#1), anchor=#1+180, shift=(#1:#2)
    }
}
\scalebox{0.75}{
 \begin{tikzpicture}[scale=1]
  \node [draw=none] (v1) at (0:1)   {};
  \node [draw=none] (v3) at (120:1)  { };
  \node [draw=none] (v4) at (240:1)  { };
  \node [draw=none,position=-80:1 from v1] (v5) { };
  \node [draw=none,position=-140:1 from v1] (v6) { };
  \node [draw=none,position=30:1 from v1] (v7) {};
  \node [draw=none,position=90:1 from v1] (v8) { };
  \node [draw=none,position=90:1 from v3] (v9) {};
  \node [draw=none,position=150:1 from v3] (v10) { };
  \node [draw=none,position=220:1 from v4] (v11) { };
  \node [draw=none,position=280:1 from v4] (v12) { };
  \node [draw=none,position=20:1 from v9] (v13)  { };
  \node [draw=none,position=90:1 from v9] (v14)  { };
  \node [draw=none,position=100:1 from v10] (v15)  { };
  \node [draw=none,position=160:1 from v10] (v16)  { };
  \node [draw=none,position=-120:1 from v10] (v17)  { };
  \node [draw=none,position=-60:1 from v10] (v18)  { };
  \node [draw=none,position=60:1.75 from v1] (v19)  { };
\hyperedge[thick,red,fill=red!20]{v1}{v3}{v4}
\hyperedge[thick,black,fill=gray!20]{v1}{v5}{v6}
\hyperedge[thick,black,fill=gray!20]{v1}{v7}{v8}
\hyperedge[thick,black,fill=gray!20]{v3}{v9}{v10}
\hyperedge[thick,black,fill=gray!20]{v4}{v11}{v12}
\hyperedge[thick,red,fill=red!20]{v9}{v13}{v14}
\hyperedge[thick,red,fill=red!20]{v10}{v15}{v16}
\hyperedge[thick,black,fill=gray!20]{v10}{v17}{v18}
\hyperedge[thick,red,fill=red!20]{v19}{v7}{v8}
\hyperedge[thick,red,fill=red!20]{v5}{v6}{v12}
\hyperedge[thick,red,fill=red!20]{v17}{v18}{v11}
\hyperedge[thick,black,fill=gray!20]{v19}{v8}{v13}
\hyperedge[thick,black,fill=gray!20]{v15}{v9}{v14}
\hyperedge[thick,black,fill=gray!20]{v3}{v8}{v9}
\end{tikzpicture}\qquad
 \begin{tikzpicture}[scale=1]
  \node [draw=none] (v1) at (0:1)   { };
  \node [draw=none] (v3) at (120:1)  { };
  \node [draw=none] (v4) at (240:1)  { };
  \node [draw=none,position=-80:1 from v1] (v5) { };
  \node [draw=none,position=-140:1 from v1] (v6) { };
  \node [draw=none,position=30:1 from v1] (v7) { };
  \node [draw=none,position=90:1 from v1] (v8) { };
  \node [draw=none,position=90:1 from v3] (v9) { };
  \node [draw=none,position=150:1 from v3] (v10) { };
  \node [draw=none,position=220:1 from v4] (v11) { };
  \node [draw=none,position=280:1 from v4] (v12) { };
  \node [draw=none,position=20:1 from v9] (v13)  { };
  \node [draw=none,position=90:1 from v9] (v14)  { };
  \node [draw=none,position=100:1 from v10] (v15)  { };
  \node [draw=none,position=160:1 from v10] (v16)  { };
  \node [draw=none,position=-120:1 from v10] (v17)  { };
  \node [draw=none,position=-60:1 from v10] (v18)  { };
  \node [draw=none,position=60:1.75 from v1] (v19)  { };
\hyperedgeF[thick,red,fill=red!20]{v1}{v3}{v4}
\hyperedgeF[thick,black,fill=gray!20]{v1}{v5}{v6}
\hyperedgeF[thick,black,fill=gray!20]{v1}{v7}{v8}
\hyperedgeF[thick,black,fill=gray!20]{v3}{v9}{v10}
\hyperedgeF[thick,black,fill=gray!20]{v4}{v11}{v12}
\hyperedgeF[thick,red,fill=red!20]{v9}{v13}{v14}
\hyperedgeF[thick,red,fill=red!20]{v10}{v15}{v16}
\hyperedgeF[thick,black,fill=gray!20]{v10}{v17}{v18}
\hyperedgeF[thick,red,fill=red!20]{v19}{v7}{v8}
\hyperedgeF[thick,red,fill=red!20]{v5}{v6}{v12}
\hyperedgeF[thick,red,fill=red!20]{v17}{v18}{v11}
\hyperedgeF[thick,black,fill=gray!20]{v19}{v8}{v13}
\hyperedgeF[thick,black,fill=gray!20]{v15}{v9}{v14}
\hyperedgeF[thick,black,fill=gray!20]{v3}{v8}{v9}
\end{tikzpicture}}}
    \end{center}
    \caption{\textit{Left.} A pictorial representation of a random $3$-hypergraph with an example of matching (in red) on it. \textit{Right}. Pictorial representation of the corresponding factor graph, where variable nodes (circle) correspond to hyperedges and function nodes (squares) correspond to nodes of the original graph.}
    \label{fig:hypergraph}
\end{figure}

\section{The planted ensemble and the inference problem}\label{sec:ens}
The inference problem we consider is given on an ensemble of (weighted) random hypergraphs which generalises the ensemble of weighted graphs discussed in \cite{SSZ_2020,SZ_2021}. This ensemble, which we denote $\mathcal H_{k,c}^N[\hat p,p]$, uses as input the coordination $k$ of the hyperedges, an integer $N\in\N$, two absolutely continuous probability densities $p$ and $\hat p$, and a real number $c\in\R^+$. A hypergraph $\mathcal G_0$ belonging to this ensemble has a set of $kN$ vertices $\mathcal V_0$ with average coordination $c+1$, and it is constructed as follows:
\begin{enumerate}
    \item A partition of the $kN$ vertices in $N$ sets of unordered $k$-plets is chosen uniformly amongst all possible partitions of the vertex set in subsets of $k$ elements. Each $k$-plet in the partition is then connected by a $k$-hyperedge, which we will call \textit{planted}. We denote $\mathcal M_0$ the set of planted hyperedges. Each planted hyperedge $e\in\mathcal M_0$ is associated to a weight $w_e$, extracted with probability density $\hat p$, independently from all the others.
    
    \item Each one of the ${{Nk}\choose{k}}-N$ remaining possible $k$-plets of vertices not in $\mathcal M_0$ is joined by a hyperedge $e$ with probability $c(k-1)!(kN)^{1-k}$. We will say that these hyperedges are \textit{non-planted} and we will denote $\mathcal E_0^{\rm np}$ their set, so that $\mathcal E_0=\mathcal M_0\cup\mathcal E_0^{\rm np}\subseteq\mathcal V_0^{\otimes k}$ is the set of all hyperedges of $\mathcal G_0$. Each non-planted edge $e\in\mathcal E_0^{\rm np}$ is associated to a weight $w_e$, extracted with probability density $p$, independently from all the others.
\end{enumerate}

By construction, the number of non-planted hyperedges will concentrate around its average $cN$ for $N\to+\infty$, so that each node has degree $1+\sZ_0$, where $\sZ_0$ is a Poissonian variable of mean $c$, $\sZ_0\sim\mathrm{Poiss}(c)$. This construction straightforwardly generalises the usual rule for generating Erd\H{o}s--R\'enyi random graphs to the case of hypergraphs. The probability of observing a certain graph $\mathcal G_0\equiv(\mathcal V_0,\mathcal E_0,\bw_0)$, with $\bw_0\coloneqq(w_e)_{e\in\mathcal E_0}$ an array of hyperedge weights, conditioned to a given set $\mathcal M_0$, is then
\begin{multline}\comprimi
\mathbb P[\mathcal G_0|\mathcal M_0]=\mathbb I(\mathcal M_0\subseteq\mathcal E_0)\prod_{{e\in\mathcal M_0}}\hat p(w_e)\prod_{{e\in\mathcal E^{\rm np}_0}}p(w_e)\\\times \left(\frac{c(k-1)!}{(kN)^{k-1}}\right)^{|\mathcal E_0^{\rm np}|}\left(1-\frac{c(k-1)!}{(kN)^{k-1}}\right)^{{{kN}\choose{k}}-|\mathcal E_0|}
\end{multline}
where $\mathbb I(\bullet)$ is the indicator function, equal to one when its argument is true, zero otherwise. By applying Bayes theorem, and using the fact that $\mathbb P[\mathcal M_0]$ is independent on $\mathcal M_0$ being uniform over all possible partitions,
\begin{equation}
\mathbb P[\mathcal M_0|\mathcal G_0]=
\mathbb P[\mathcal G_0|\mathcal M_0]\frac{\mathbb P[\mathcal M_0]}{\mathbb P[\mathcal G_0]}\propto 
\mathbb P[\mathcal G_0|\mathcal M_0].
\end{equation}
We parametrise the posterior by associating to each matching $\mathcal M_0$ the \textit{matching map} $\bfm\colon\mathcal E_0\to \{0,1\}^{|\mathcal E_0|}$ such that $m_e=\mathbb I(e\in\mathcal M_0)$. Note that a matching map satisfies the constraint $\sum_{e\in\partial v}m_e=1$ for each $v\in\mathcal V$, where $\partial v$ is the set of hyperedges that are incident to $v$. It is clear that there is a one-to-one correspondence between a matching $\mathcal M_0$ and its map $\bfm$: by an abuse of notation, we will therefore use $\mathcal M_0$ and its map $\bfm$ interchangeably, and write $\mathbb P[\mathcal M_0|\mathcal G_0]\equiv \mathbb P[\bfm|\mathcal G_0]$. We denote in particular $\bfm^\star$ the matching map corresponding to ground truth, i.e., the planted matching. Our goal is to use the posterior to produce an estimator $\hat \bfm$ of $\bfm^\star$. As in the $k=2$ case \cite{SSZ_2020}, the estimator can be chosen in such a way that a certain measure of distance from the true planted matching $\bfm^\star$ is minimised. A possible measure of distance is the function
\begin{equation}\label{errore}
\varrho(\bfm)\coloneqq\frac{1}{2N}
\sum_{e}\mathbb I(m_e\neq m_e^\star).
\end{equation}
The estimator minimising the quantity above can be constructed by minimising the expectation of each element of the sum over the posterior, i.e., choosing for each edge $e$ of the graph
\begin{equation}
m_e^{\rm s}\coloneqq \arg\max_{m\in\{0,1\}}\mathbb P[m_e=m|\mathcal G_0]
\end{equation}
where $\mathbb P[m_e=m|\mathcal G_0]$ is the marginal probability of $m_e$, value of the matching map on the edge $e$. We call this estimator \textit{symbol maximal a posteriori} (sMAP), following the nomenclature adopted in the study of error correcting codes \cite{error_corr_codes}. However, by construction, the estimator $\hat\bfm$ \textit{is not} a matching map in general. A different estimator, which instead provides a genuine matching, can be obtained considering
\begin{equation}
\bfm^{\rm b}\coloneqq \arg\max_{\bfm \text{ matching}}\mathbb P[\bfm|\mathcal G_0],
\end{equation}
called \textit{block maximal a posterior} (bMAP) estimator. The bMAP minimises $\varrho(\bfm)$ over the space of matching maps and is therefore a matching map. In what follows, we will study $\mathbb E[\varrho]$ for both the sMAP and the bMAP, the average $\mathbb E[\bullet]$ to be intended over the ensemble $\mathcal H_{k,c}^N[\hat p,p]$ for $N\to+\infty$.

\section{Belief propagation algorithm}\label{sec:bp}
\subsection{A preliminary pruning of ${\mathcal G_0}$}
As in the $k=2$ case, if the distributions $p$ and $\hat p$ have different support, it will be possible to identify some hyperedges as planted or non-planted simply by direct inspection. Assuming $\Gamma\coloneqq\mathrm{supp}(p)\cap\mathrm{supp}(\hat p)$ to be of nonzero Lebesgue measure, it is clear that if an edge $e$ has $w_e\in\mathrm{supp}(p)\setminus\Gamma$, then $m^\star_e=0$. Similarly, if $w_e\in\mathrm{supp}(\hat p)\setminus\Gamma$, then $m_e^\star=1$. By consequence, a preliminary {pruning} of the graph is possible by removing all edges that are immediately identifiable \footnote{Note, in particular, that if $e\in\mathcal E_0$ is identified as planted, it must be removed alongside its endpoints and all hyperedges attached to them}. Let us define the portions of mass of the two distributions over $\Gamma$ as $\mu\coloneqq\int_\Gamma p(w)\dd w$ and $\hat\mu\coloneqq\int_\Gamma \hat p(w)\dd w$, so that, after such pruning, $w_e\sim \hat P(w)\coloneqq{\hat \mu}^{-1}\hat p(w)\mathbb I(w\in\Gamma)$ if $e\in\mathcal M_0$ and $w_e\sim P(w)\coloneqq{\mu}^{-1} p(w)\mathbb I(w\in\Gamma)$ if $e\not\in\mathcal M_0$. The pruned hypergraph, that we will call $\mathcal G_1=(\mathcal V_1,\mathcal E_1,\bw_1)$, has $\mathcal V_1\subseteq\mathcal V_0$, $\mathcal E_1\subseteq\mathcal E_0$ and $w_e\in\Gamma$ for all edges $e\in\mathcal E_1$. Moreover, $|\mathcal V_1|=kN\hat\mu$, each node having one incident planted hyperedge and $\sZ_1$ incident non-planted hyperedges, with $\sZ_1\sim\mathrm{Poiss}(\gamma)$, $\gamma\coloneqq c\mu\hat \mu^{k-1}$ \footnote{Each non-planted hyperedge $e$ will survive with probability $\mu\hat\mu^k$, as both $e$ and the planted hyperedges incident at its endpoints have to survive; however, as anticipated, after the pruning $|\mathcal V_1|=\hat\mu kN$.}. Finally, let us call $\mathcal M_1\coloneqq\{e\in\mathcal M_0\mid w_e\in\Gamma\}$.

Once the graph $\mathcal G_1$ has been obtained, an additional, elementary observation can further reduce the size of the problem. Due to the fact that $\mathbb P[\sZ_1=0]=\e^{-\gamma}\neq 0$ at finite $c$, for large $N$ the graph will contain leaves with finite probability. For each of these leaves, the single incident hyperedge $e$ can be classified as an element of $\mathcal M_1$, and removed from the graph alongside with its endpoints and their corresponding incident hyperedges. In this way, we can proceed recursively in a new pruning of $\mathcal G_1$ until a new hypergraph $\mathcal G=(\mathcal V,\mathcal E,\bw)$ is obtained that cannot be further pruned. This graph has no leaves by construction and all the edges $e\in\mathcal E$ have $w_e\in\Gamma$. 

To compute the fraction of surviving hyperedges, let us consider the graph $\mathcal G_1$ and an edge $e\in\mathcal M_1$. We denote $1-\hat q$ the probability that one of the endpoints of $e$ is a leaf at a certain point of the second pruning: if this is the case, $e$ will be pruned. Similarly, if $e\in\mathcal E_1\setminus\mathcal M_1$ is non-planted, we denote $1-q$ the corresponding probability that one of its endpoints will become a leaf at some point. The quantity $\hat q$ satisfies the equation
\begin{subequations}\label{qqhat}
\begin{equation}
1-\hat q=\sum_{n=0}^\infty\frac{\e^{-\gamma}}{n!}\left[(1-q^{k-1})\gamma\right]^n=\e^{-\gamma q^{k-1}}, \label{eq:qhat}
\end{equation}
as it is sufficient, for each non-planted edge incident to a given vertex to be pruned, that \textit{one} of the remaining $k-1$ endpoints requires pruning. The equation for $q$ is simpler as an endpoint of a non-planted hyperedge is removed if, and only if, its incident planted hyperedge is pruned, therefore
\begin{equation}
q=\hat q^{k-1}. 
\end{equation}
\end{subequations}
We numerically verified Eqs.~\eqref{qqhat} in Appendix \ref{app:sparse}. As a result, a node of $\mathcal G$ has coordination $1+\sZ$, where $\sZ$ is a zero-truncated Poisson distribution of parameter $q^{k-1}\gamma$, $\sZ\sim\mathrm{ZTPoiss}(q^{k-1}\gamma)$ \footnote{Given a node $v$ with $\sZ_1\sim\mathrm{Poiss}(\gamma)$ non-planted edges, each of them will be present with probability $q^{k-1}$, so that the probability that $v$ has $z$ surviving edges is $\mathbb P[\sZ=z]=\sum_{n=z}^\infty \frac{\gamma^n}{n!}\e^{-\gamma}\binom{n}{z}q^{(k-1)z}(1-q^{k-1})^{n-z}=\frac{1}{z!}\e^{-\gamma q^{k-1}}(q^{k-1}\gamma)^z$. As vertices with $\sZ=0$ are removed from the graph, the resulting distribution is therefore $\mathrm{ZTPoiss}(q^{k-1}\gamma)$.}. We will denote $\mathcal M$ the set of unidentified planted hyperedges, and fix $m_e\equiv m_e^\star=1$ for all the identified hyperedges $e\in\mathcal M_0\setminus\mathcal M$.

\subsection{Back to the posterior and Bayes-optimality}

At this point, we have exploited the information deriving from the the weights and the topology separately. To further proceed in the estimation of $\bfm^\star$, the optimal approach goes through the calculation of the posterior
\begin{equation}
\mathbb P[\bfm|\mathcal G]\propto\prod_{e\in\mathcal M}\left[\frac{\hat P(w_e)}{P(w_e)}\right]^{m_e}\prod_{v\in\mathcal V}\mathbb I\left(\sum_{e\in\partial v}m_e=1\right),\label{post}
\end{equation}
where the requirement that $\bfm$ is a matching map is explicitly enforced by the indicator function. Estimating the measure in Eq.~\eqref{post} is pivotal to obtain both the bMAP and the sMAP. To do so, we consider
\begin{multline}\label{post2}\comprimi
\nu_\beta(\bfm)\propto\exp\left(-\beta\sum_e m_e\omega_e\right)\prod_{v\in\mathcal V}\mathbb I\left(\sum_{e\in\partial v}m_e= 1\right),
\end{multline}
where we have denoted
\begin{equation}
\omega_e\coloneqq-\ln\frac{\hat P(w_e)}{P(w_e)}\quad \forall e\in\mathcal E,    
\end{equation}
and we have introduced a new parameter $\beta>0$ (hence the change of notation). The parameter is such that, for $\beta=1$, Eq.~\eqref{post2} corresponds to Eq.~\eqref{post}: this means that, by sampling from $\nu_1$, we sample from the correct posterior and we are in a Bayes optimal setting that leads to the lowest possible error $\rho$. Given a real function $f(\bfm_1,\bfm_2)$ of two matching maps,  assuming that $\bfm_1$, $\bfm_2$ and $\bfm$ are independent samples from $\nu_1$, then $\mathbb E[f(\bfm,\bfm^\star)]=\mathbb E[f(\bfm_1,\bfm_2)]$, a property known in physical jargon as Nishimori condition \cite{nishimori1980}. Importantly, validity of the Nishimori condition implies the absence of replica symmetry breaking.  

On the other hand, $\arg\max_{\bfm}\nu_1(\bfm)$ can be obtained as the support of $\nu_\beta$ in the limit $\beta\to+\infty$.

\subsection{Belief-propagation equations}
Due to the sparse nature of the hypergraphs under study, a natural tool to estimate the posterior of the problem is \textit{belief propagation} \cite{mezard2009}. The belief propagation equations for the minimum-weight matching problem on hypergraphs, or multi-index matching problem (MIMP), are derived in Ref.~\cite{Martin2004,MMR_2005}. The algorithm runs on a \textit{factor graph} obtained from the original weighted hypergraph representing each hyperedge $e$ by a variable node, and each vertex $v\in\mathcal V$ by a factor node. Variable nodes correspond to the variables $m_e$ and are associated to a weight $\e^{-\beta m_e\omega_e}$, $e\in\mathcal E$; each factor node, on the other hand, represents the local constraint $\sum_{e\in\partial v}m_e=1$, $v\in\mathcal V$, see Fig.~\ref{fig:hypergraph}. The analysis of our case follows straightforwardly the study of the minimum-weight MIMP \cite{Martin2004,MMR_2005}, the main (but crucial, in the statistical analysis) difference being the fact that the weights have in our case the meaning of log-likelihood on differently distributed weights. For each edge $(e,v)$ of the factor graph --- joining the variable node $e$ corresponding to the hyperedge $e\in\mathcal E$ with the factor node $v$ corresponding to the node $v\in\mathcal V$ --- we introduce two ``messages'', namely
\begin{subequations}
\begin{equation}\comprimi
\hat \nu_{v\to e}(m)\propto
 \sum_{\mathclap{\{m_{\tilde e}\}_{\tilde e\in\partial v\setminus e}}}\quad\mathbb I\Big(m+\sum_{\mathclap{\tilde e\in\partial v\setminus e}}m_{\tilde e}= 1\Big)
\prod_{\mathclap{\tilde e\in\partial v\setminus e}} \nu_{\tilde e\to v}(m_{\tilde e})
\end{equation}
and
\begin{equation}
\nu_{e\to v}(m)\propto\e^{-\beta m\omega_e} \prod_{\mathclap{u\in\partial e\setminus v}}\hat \nu_{u\to e}(m),
\end{equation}\label{bp1}\end{subequations}
where $\partial e$ is the set of endpoints of $e$. The message $\nu_{e\to v}$ mimics the marginal probability of the variable $m_e$ in absence of the endpoint $v$. The equations are obtained in the hypothesis of a tree-like structure of the factor graph, so that the incoming contributions in a node can be considered independent. Exploiting $m$ being a binary variable, it is convenient to parametrise both marginals by means of \textit{cavity fields}, namely write
\begin{equation}
\hat\nu_{v\to e}(m)\eqqcolon\frac{\e^{\beta m h_{v\to e}}}{1+\e^{\beta h_{v\to e}}},\quad \nu_{e\to v}(m)\eqqcolon\frac{\e^{\beta m \eta_{e\to v}}}{1+\e^{\beta \eta_{e\to v}}},
\end{equation}
so that the belief propagation equations in Eq.~\eqref{bp1} become
\begin{subequations}\label{bp2}
\begin{align}
    h_{v \to e} &= - \frac{1}{\beta} \ln\Big[
    \sum_{\mathclap{\tilde{e} \in \partial v \setminus e}}\e^{\beta ( \eta_{\Tilde{e} \to v} - \omega_{\Tilde{e}} )}\Big],\\
    \eta_{e \to v} &= \sum_{\mathclap{u \in \partial e\setminus v}} h_{u \to e}.
\end{align}
\end{subequations}
Such equations specify a \textit{belief propagation algorithm} (BPA) to estimate the marginals of the posterior probability: we will use this algorithm, which is exact if the factor graph is a tree, to estimate the marginals of the true posterior. In particular, the marginal distribution of the variable $m_e$ corresponding to the hyperedge $e\in\mathcal E$ is obtained as
\begin{equation}
\begin{split}
    \nu_e(m)&\propto \e^{-\beta m_e\omega_e}\prod_{v\in\partial e}\hat\nu_{v\to e}(m)\\&\propto \exp\Big[\beta m_e\Big(\sum_{v\in\partial e}h_{v\to e}-\omega_e\Big)\Big].
\end{split}
\end{equation}
A hyperedge $e$ can be therefore selected if $\nu_e(1)\geq\sfrac{1}{2}$. In other words, we can construct $\hat\bfm\equiv\bfm^{\rm s}$ ($\hat\bfm\equiv\bfm^{\rm b}$, respectively) computing the fields $h_{v\to e}$ for $\beta=1$ ($\beta\to+\infty$, respectively) and then taking
\begin{equation}
 \hat m_e=\theta\Big(\sum_{v\in\partial e}h_{v\to e}-\omega_e\Big).\label{nue1}
\end{equation}

\begin{figure*}
    \subfloat[Average error $\ln{\mathbb E[\varrho]}$ for $k=2$. The transition from partial to full recovery is continuous. For $\beta\to+\infty$ the transition takes place at $\lambda=4$ \cite{MMX_2019} and it is proven to be of infinite order \cite{SSZ_2020,ding2021}. By consequence, a full recovery phase exists for $\lambda>4$ at the Bayes optimal value $\beta=1$. Note however that this does not hold for all values of $\beta$. The RS ansatz is proven to be the correct one for $\beta\to+\infty$ \cite{Bayati2011} and must be correct for $\beta=1$ due to the Nishimori conditions. The BP algorithm is indeed found to converge correctly for all values of $\beta$. \label{fig:phase2}]{\centering\includegraphics[height=0.3\textwidth]{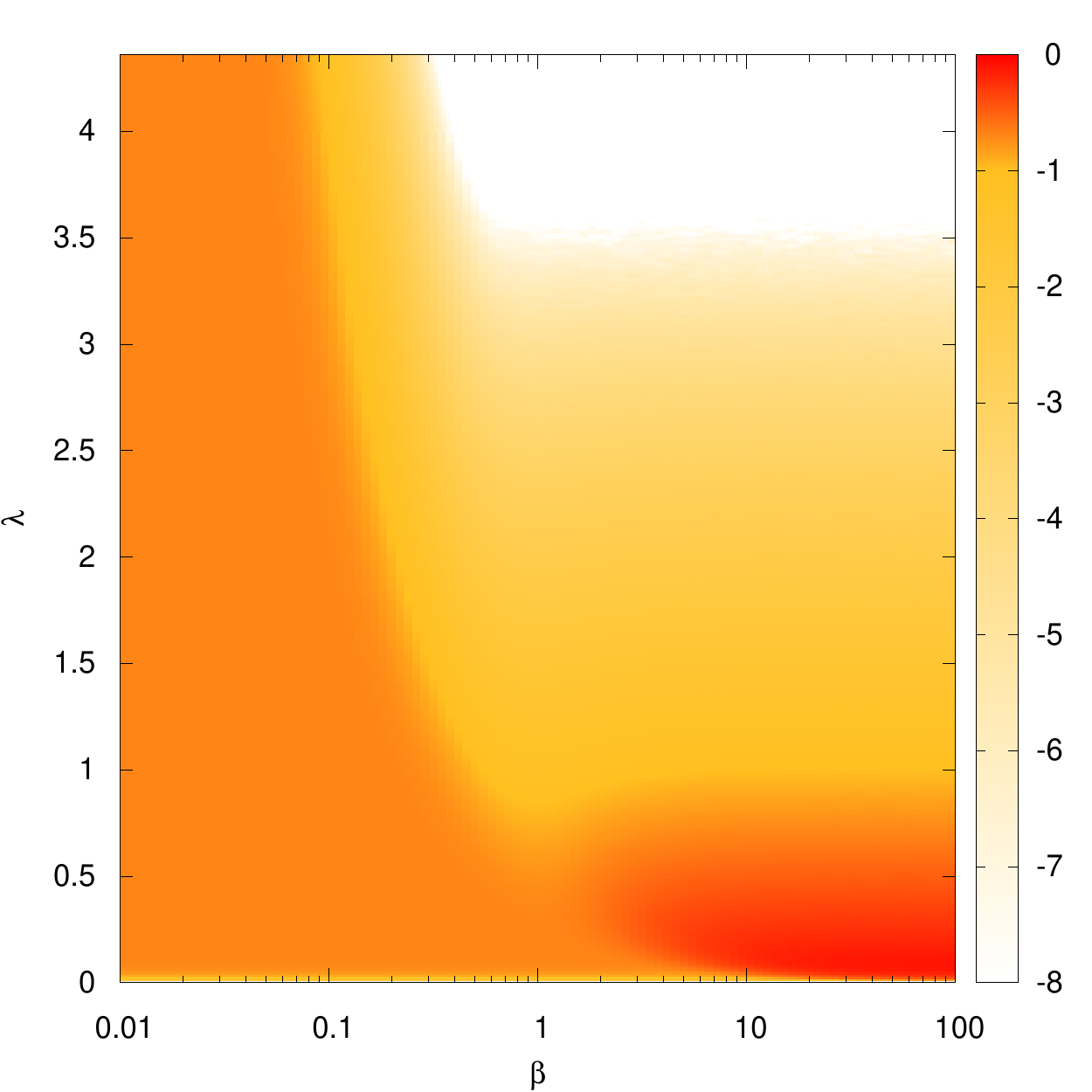}}\qquad
    \subfloat[Average error $\ln{\mathbb E[\varrho]}$ achievable by a BPA for $k=3$ as predicted by the PDA. The sharp color change when approaching the full recovery region (in white) is due to the discontinuous nature of the transition taking place on the continuous line representing $\lambda_{\rm alg}$. In the region above the dashed line the partial recovery solution is thermodynamically unstable at that value of $\beta$. Perfect recovery is information theoretically possible above the dashed line at $\beta=1$, i.e. above $\lambda_{\rm it}=0.43(1)$ and impossible below. Note that the results are obtained in the RS assumption which is correct at $\beta=1$ but needs to be verified for $\beta\neq 1$.  \label{fig:phase3}]{\centering\includegraphics[height=0.3\textwidth]{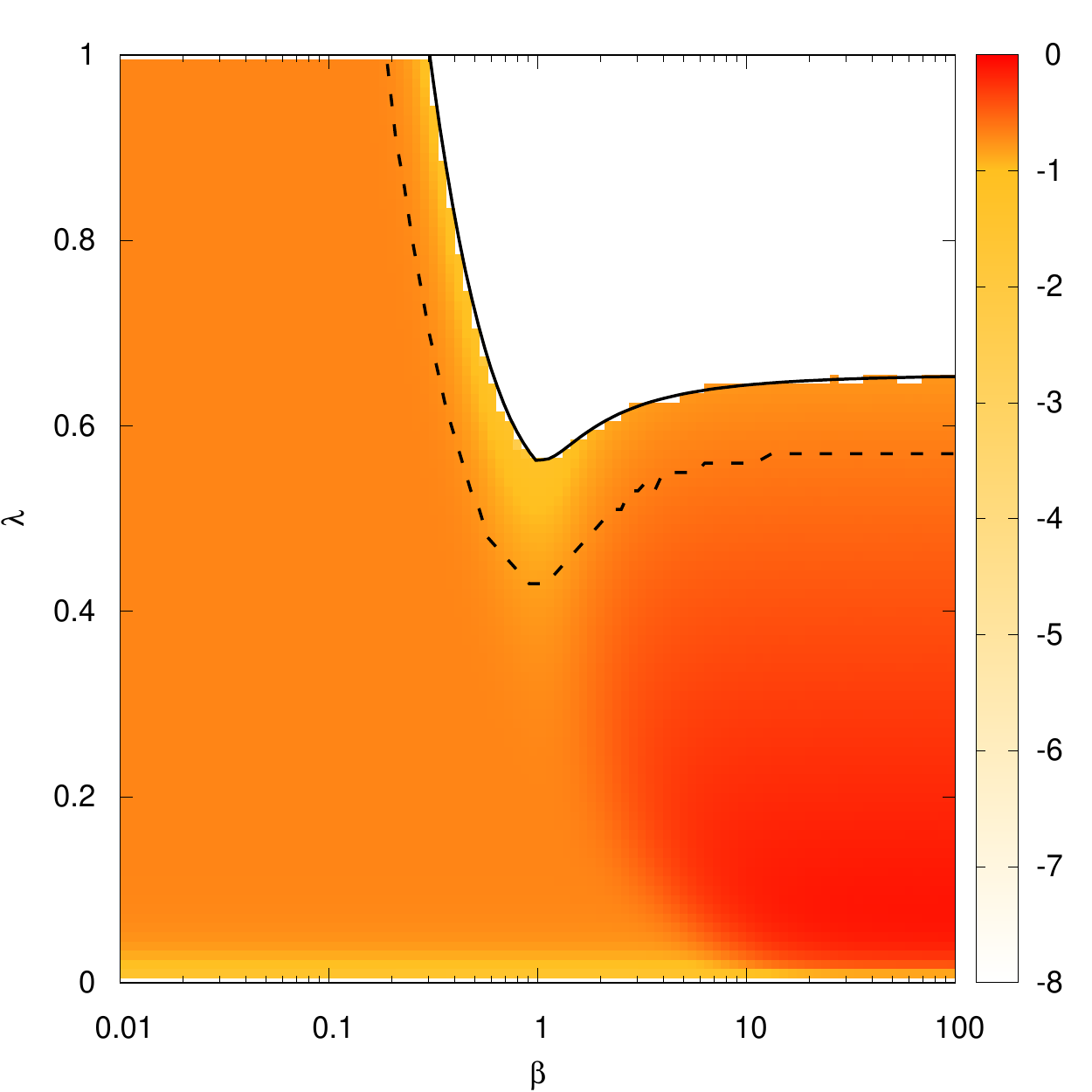}}\qquad
    \subfloat[Average error ${\ln \mathbb E[\varrho]}$ achieved by a BPA in the Bayes-optimal setting ($\beta=1$) for the planted $(2+3)$-MIMP. Here $r=0$ corresponds to the pure $k=2$ case, whilst $r=1$ corresponds to the pure $k=3$ case. The first-order algorithmic transition $\lambda_{\rm alg}$ (continuous line) becomes of second order at $r=0.244(4)$ (full dot). The dashed line corresponds to the value $\lambda_{\rm IT}$ as a function of $r$. \label{fig:phaseB1}]{\centering\includegraphics[height=0.3\textwidth]{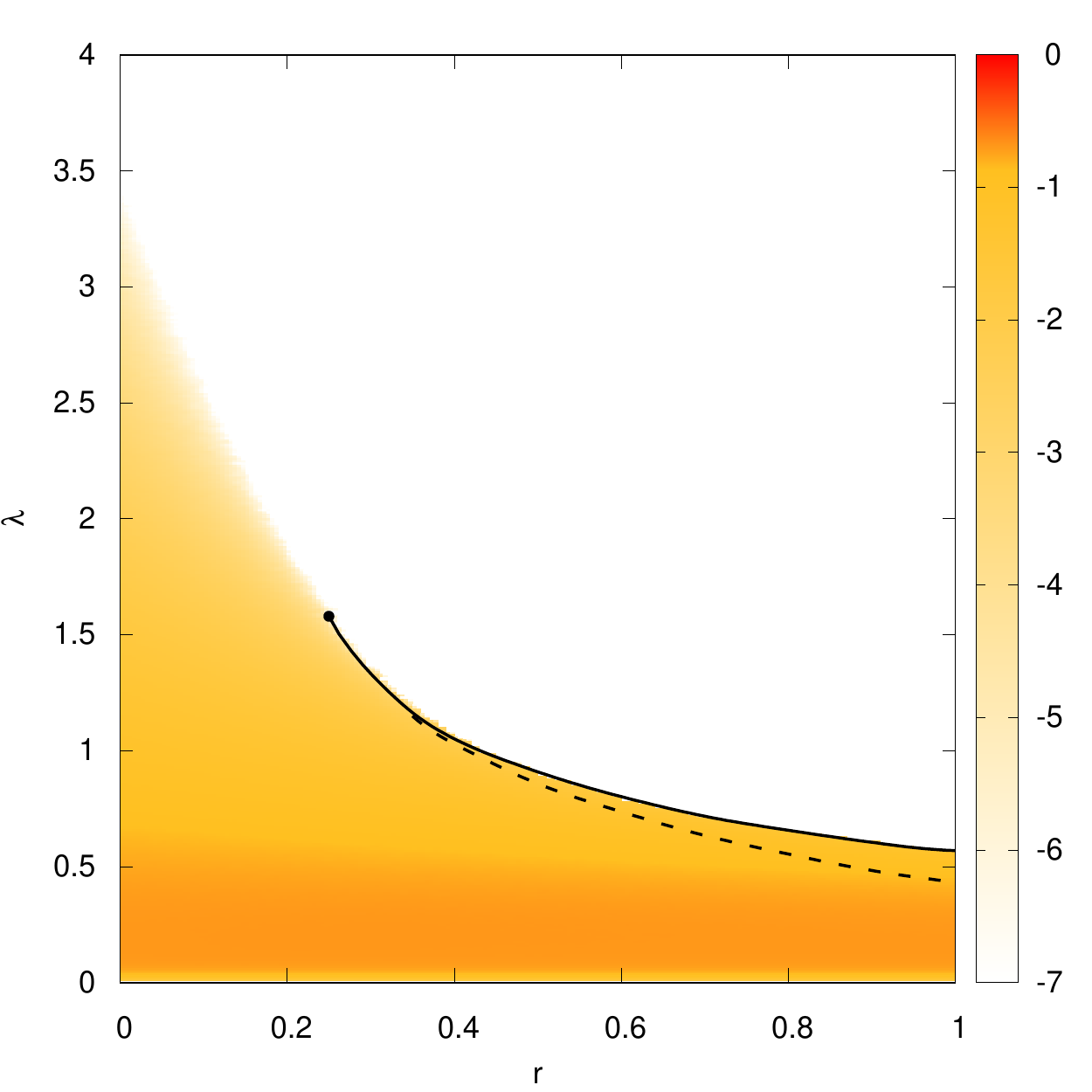}}
    \caption{Average error ${\mathbb E[\varrho]}$ for the planted MIMP obtained using a PDA with $10^4$--$10^5$ fields. For our numerical simulation, we used here $c=50$ as we observed no sensible dependence on $c$ for larger values of the average degree and $\lambda\gtrsim\sfrac{1}{c}$: note that for $\lambda\to 0$, the effect of the finite-$c$ approximation becomes evident as a full recovery region appears near the origin (see Appendix \ref{app:sparse}).}
\end{figure*}

\subsection{Recursive distributional equations}

To study the performances of the algorithm in the $N\to+\infty$ limit at any $\beta$, we can write down a set of recursive distributional equations (RDEs) involving random variables whose statistics follow the one of the cavity fields in the BPA. Following \cite{CKKVZ_2010,SSZ_2020,Martin2004}, let us introduce the random variables $\hat{\sH}$ and $\sH$ distributed as the cavity fields $h_{v\to e}$ on a planted and non-planted hyperedge, respectively. Let us also denote $\hat{\sOmega}$  and $\sOmega$ two random variables distributed as $\omega_e$ on planted and non planted $k$-hyperedges respectively. In the large-size limit, such random variables satisfy the following recursive distributional equations (RDEs)
\begin{subequations}\label{RDE1}
\begin{equation}
    \hat{\sH} \comprimi \textstyle\stackrel{\rm d}{=} -\frac{1}{\beta}\ln\left[
    \sum\limits_{v=1}^{\sZ}\exp\left(\beta\sum\limits_{u=1}^{k-1} \sH_{vu}-\beta\sOmega_v\right)\right],
\end{equation}
and
\begin{equation} 
    \sH \comprimi \stackrel{\rm d}{=}\begin{cases}
    \hat\sOmega-\sum_{u=1}^{k-1} \hat\sH_{u}     
    
    \\\qquad\text{with probability } 1-\hat q,\\ -\frac{1}{\beta}\ln\left[\exp\left(-\beta\hat\sH\right)+\exp\left(\beta\sum_{u=1}^{k-1} \hat\sH_{u}-\beta
    \hat\sOmega\right)\right]\\\qquad\text{with probability } \hat q,
    \end{cases}
\end{equation}
\end{subequations}
with $\sZ\stackrel{\rm d}{=}\mathrm{ZTPoiss}
(q^{k-1}\gamma)$. The equations above are straightforward generalisations of the $k=2$ case discussed in Ref.~\cite{SSZ_2020}. In particular, for $\beta\to+\infty$ the RDEs become
\begin{subequations}\label{RDE0}
\begin{align} 
    \hat{\sH} &\textstyle \stackrel{\rm d}{=} \min\limits_{1\leq v\leq \sZ}\left\{\sOmega_v-\sum_{u=1}^{k-1} \sH_{vu} \right\},\\
    \sH &\comprimi \stackrel{\rm d}{=}\begin{cases}\hat\sOmega-\sum_{u=1}^{k-1} \hat\sH_{u}&\text{with prob. } 1-\hat q,\\\min\left(\hat\sOmega-\sum_{u=1}^{k-1} \hat\sH_{u},\hat\sH \right)&\text{with prob. } \hat q.
    \end{cases}\label{RDE0b}
\end{align}
\end{subequations}

Due to Eq.~\eqref{errore} and Eq.~\eqref{nue1}, the average of the reconstruction error for both the bMAP and the sMAP estimator is then obtained as
\begin{equation}\comprimi
    \mathbb{E} [\varrho] = \frac{\hat{\mu}\hat q^k}{2} \mathbb{P} \left[ \sum_{v=1}^k \hat{\sH}_v \leq \hat{\sOmega} \right] + \frac{ \gamma\hat{\mu} q^k}{2} \mathbb{P} \left[ \sum_{v=1}^k \sH_v > \sOmega \right],\label{eq:err}
\end{equation}
the difference between the two cases being the chosen value of $\beta$ in the RDEs. For $\beta \to+\infty$ Eq.~\eqref{eq:err} can be further simplified (see Appendix \ref{app1}) as
\begin{equation}
    \mathbb{E} [\varrho] = \hat{\mu}\hat q^k \mathbb{P} \left[ \sum_{v=1}^k \hat{\sH}_v \leq \hat{\sOmega} \right].\label{eq:rho0}
\end{equation}
Note that, for any value of $\beta$, $k$ and $c$ and for any pair of distributions $p$ and $\hat p$, the RDEs above admit the solution $\hat\sH=-\sH=+\infty$, which corresponds to a \textit{full recovery} of the hidden signal, i.e., $\mathbb E[\varrho]=0$. 

The RDEs also allow to estimate the Bethe free energy at any $\beta$ \cite{mezard2009} in the large $N$ limit, which is defined, on a given instance $\mathcal G_0$ of $\mathcal H^N_{k,c}[\hat p,p]$, as
\begin{multline}
f^{\rm B}_{\mathcal G_0}(\beta)\coloneqq
\comprimi\frac{1}{N}\sum_{a\in\hat{\mathcal{M}}}\omega_a\\+\frac{k-1}{\beta N}\sum_{a\in\mathcal E}\ln\Big[1+\exp\Big(\beta\sum\limits_{\mathclap{u\in\partial a}}h_{u\to a}-\beta\omega_e\Big)\Big]\\\comprimi
-\frac{1}{N\beta}\sum_{v\in\mathcal V}\ln\Big[
\sum_{a\in\partial v}\exp\Big(\beta\sum\limits_{\mathclap{ u\in\partial a\setminus v}}h_{u\to a}-\beta\omega_e\Big)\Big].
\end{multline}
This quantity estimates the log-likelihood $-\frac{1}{N\beta}\ln\sum_{\bfm}\nu_\beta(\bfm)$ within the tree-like assumption. In the $N\to+\infty$ limit, $f_{\mathcal G_0}^{\rm B}(\beta)$ is expected to concentrate on
\begin{multline}
f^{\rm B}(\beta)=(1-\hat\mu\hat q^k)\mathbb E[\hat\sOmega]
\comprimi-\frac{\hat\mu\hat q^k k}{\beta}\mathbb E\ln\left(\e^{-\beta\hat\sH}+\e^{\beta\sum_{u=1}^{k-1}\hat\sH_u-\beta\hat\sOmega}\right)\\
+\frac{(k-1)\hat\mu\hat q^{k}}{\beta}\mathbb E\ln\left(1+\e^{\beta\sum_{u=1}^{k} \sH_{u}-\beta\sOmega}\right)\\
+\frac{\gamma(k-1)\hat\mu q^{k}}{\beta}\mathbb E\ln\left(1+\e^{\beta\sum_{u=1}^{k} \hat\sH_{u}-\beta\hat\sOmega}\right).
\end{multline}
For $\beta\to+\infty$, $f^{\rm B}(\beta)$ converges to (minus) the log-likelihood of the bMAP estimator,
\begin{multline}\textstyle
\lim\limits_{\beta\to+\infty}f^{\rm B}(\beta)= (1-\hat\mu\hat q^k)\mathbb E[\hat\sOmega]\\\comprimi
+\hat{\mu}\hat q^k\mathbb E\left[\hat\sOmega\theta\left(\sum_{v=1}^k \hat{\sH}_v \geq \hat{\sOmega}\right)\right]+\hat\mu\gamma q^k\mathbb E\left[\sOmega\theta\left(\sum_{v=1}^k {\sH}_v \geq {\sOmega}\right)\right], \label{eq:likelihood}  
\end{multline}
where the random variables $\sH$ and $\hat\sH$ satisfy the set of RDEs \eqref{RDE0}. Note that the Bethe free energy associated to the infinite-fields fixed point is simply $f^\star=\mathbb E[\hat\sOmega]$ and corresponds to (minus) the log-likelihood of $\bfm^\star$.

\section{The partial-full recovery transition in the planted MIMP}\label{sec:transit}
The RDEs in Eq.~\eqref{RDE1} can be solved numerically by means of a population dynamics algorithm (PDA) \cite{mezard2009}. In the numerical results presented below, the planted weights are independently generated from an exponential distribution of mean $\lambda$, $\hat p=\mathrm{Exp}(\lambda)$, whilst the non-planted edges have weights uniformly distributed on the interval $[0,c]$, $p=\mathrm{Unif}([0,c])$. We will focus on the $c\to +\infty$ limit 
(the finite-$c$ case exhibits a qualitatively similar phenomenology). 
In Fig.~\ref{fig:phase3} we present the reconstruction error achievable via a BPA predicted by the PDA for different values of $\beta$ and $\lambda$ for $k=3$. The value $\beta=1$ corresponds to the error associated to the sMAP estimated via a BPA, whereas the bMAP is obtained for $\beta\to+\infty$. The figure makes evident that, at given $\lambda$, the performances at $\beta=1$ are optimal. We see that there is a sharp transition between a region with $\mathbb E[\varrho]>0$ and a region with $\mathbb E[\varrho]=0$. For $k=2$ a similar phase diagram can be drawn, see Fig.~\ref{fig:phase2}: the nature of the transition, however, is different. The transition towards the full recovery phase is continuous and it has been proven that it is of infinite order as $\beta\to+\infty$ \cite{SSZ_2020,ding2021}. 
In Fig.~\ref{fig:phase3} we also present by a dashed line the value of $\lambda$ above which the partial recovery solution is thermodynamically unstable, or in other words metastable. This line is computed by comparing the Bethe free energy of the partial recovery fixed point to the fixed point corresponding to the planted solution.

Let us focus now on the $\beta=1$ line and on the $\beta\to+\infty$ line, corresponding to the estimation via a BPA of the sMAP and the bMAP respectively.

\begin{figure}[t]
    \centering
   \includegraphics[width=\columnwidth]{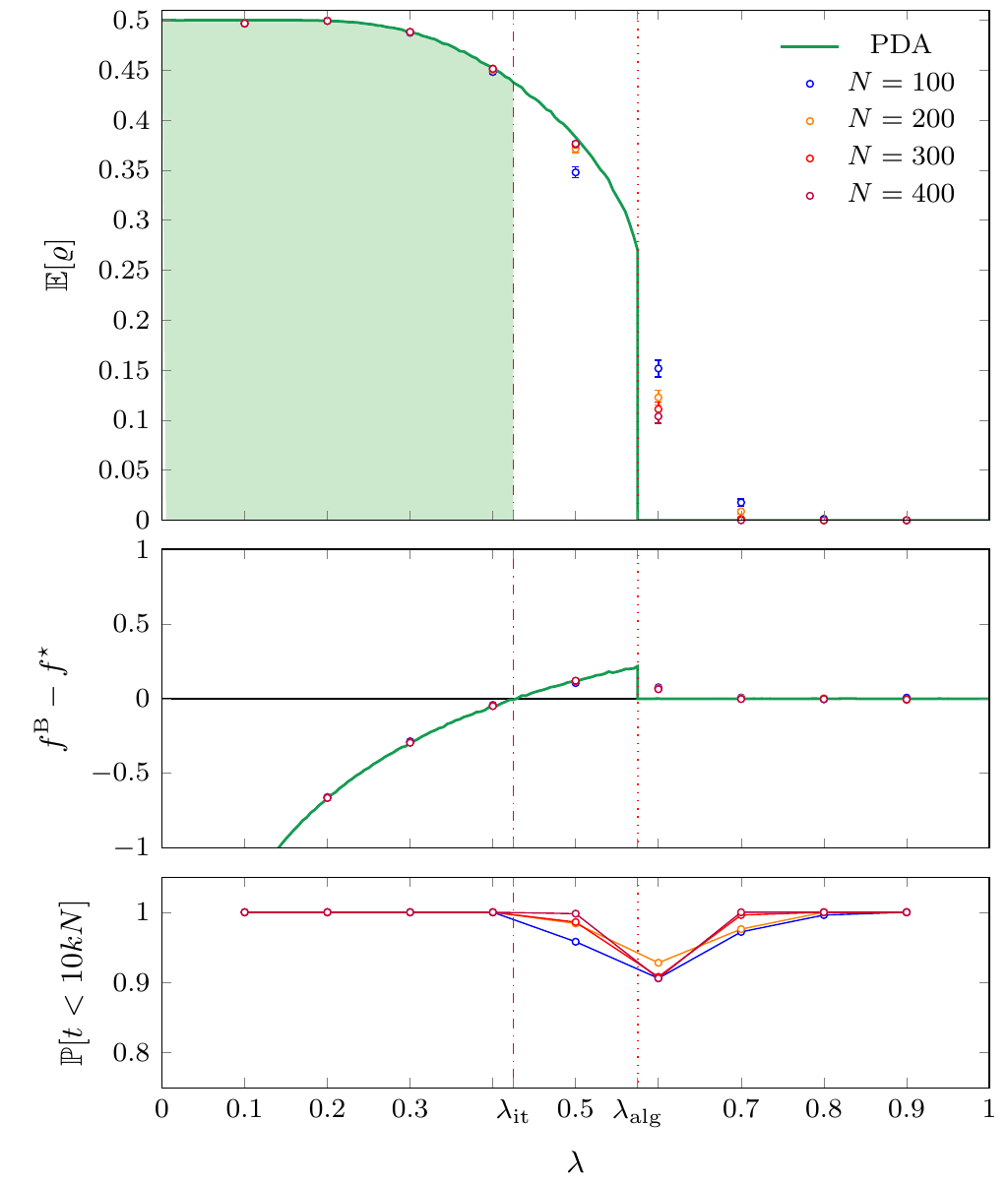}
    \caption{Numerical results for the planted MIMP at $\beta=1$. Smooth curves are obtained from a PDA solving the RDEs in Eq.~\eqref{RDE1} at $\beta=1$ with $k=3$. We assume that $\hat p=\mathrm{Exp}(\lambda)$ and $p=\mathrm{Unif}([0,c])$. The PDA used a population of $10^5$ fields updated $200$ times for each value $\lambda$. For our population dynamics numerics, we assumed here $c=300$, to reduce as much as possible the finite-$c$ effects near the origin (see Appendix \ref{app:sparse}). Dots are obtained by running a BPA on $5\cdot10^2$ instances of the ensemble $\mathcal H_{3,50}^N[\hat p,p]$. \textit{Top.} Average error $\mathbb E[\varrho]$ obtained via a BPA at $\beta=1$: the PDA prediction is compared with the results of numerical simulations. \textit{Center.} Difference between the Bethe free energy obtained via the PDA and the free energy of the planted solution. The fixed point obtained by the PDA is thermodynamically unstable for $\lambda_{\rm it}<\lambda<\lambda_{\rm alg}$. \textit{Bottom.} Probability $\mathbb P[t<10kN]$ that the algorithm requires a number of sweep smaller than $10kN$ to reach convergence: particularly hard instances appear for $\lambda_{\rm it}<\lambda<\lambda_{\rm alg}$, where we estimate $\mathbb P[t<10kN]<1$.
    }
    \label{fig:rhoexp1}
\end{figure}

\begin{figure}[t]
    \centering
   \includegraphics[width=\columnwidth]{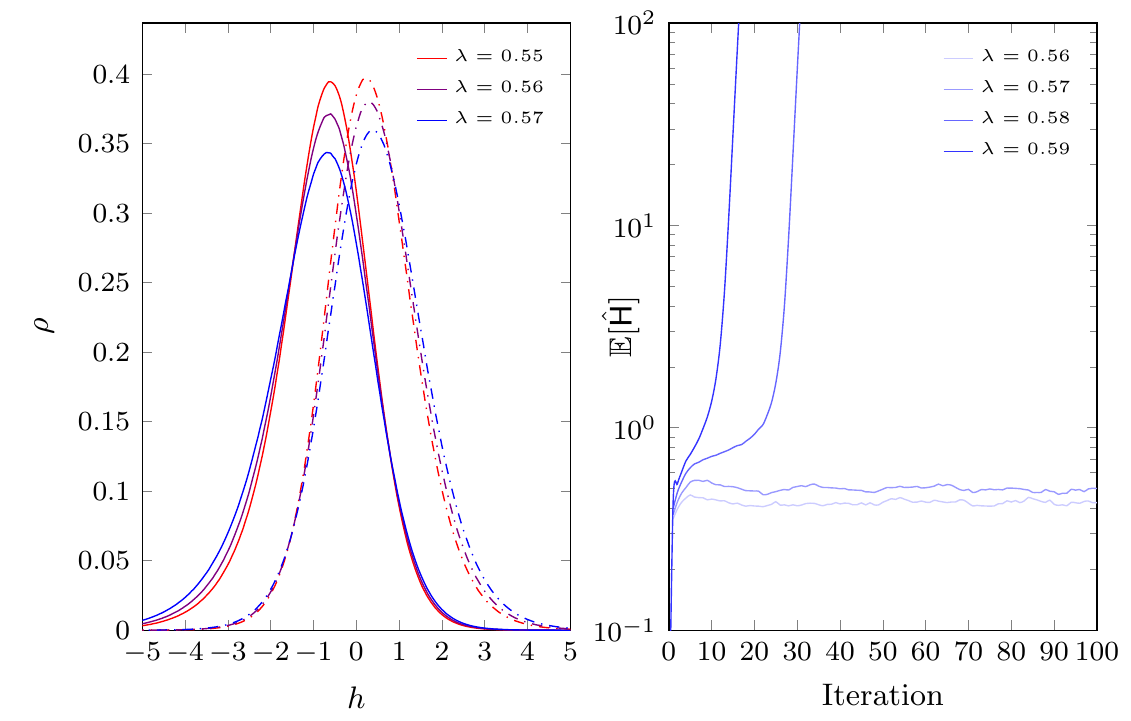}
    \caption{\textit{Left.} Distribution of the cavity fields on the planted (continuous line) and nonplanted (dotted line) hyperedges for $k=3$ and different values of $\lambda$ near the transition point. \textit{Right.} Value of $\mathbb E[\hat\sH]$ as function of the iteration step $t$ in the PDA at $\beta=1$ with $k=3$ and $c=100$, a value large enough to see no dependence on $c$ in our results in the considered range of $\lambda$: a sharp change of behavior is observed at $\lambda_{\rm alg}=0.578(1)$.
    }
    \label{fig:disth}
\end{figure}

\paragraph{The sMAP estimator} In Fig.~\ref{fig:rhoexp1} we present the results obtained by solving the RDEs in Eq.~\eqref{RDE1} with $\beta=1$ by means of a PDA, and by estimating $\mathbb E[\varrho]$ for different values of $\lambda$.
As anticipated, the phenomenology is different from the $k=2$ case, where a continuous transition at $\lambda_{\rm alg}\simeq 4$ is observed \cite{SSZ_2020}: for $k=3$, a sharp jump in $\mathbb E[\varrho]$ takes place at $\lambda_{\rm alg}=0.578(1)$, so that $\mathbb E[\varrho]=0$ for $\lambda>\lambda_{\rm alg}$, i.e., perfect recovery of the planted configuration is achieved, and the solution $\hat\sH=-\sH=+\infty$ is found with belief propagation. For $\lambda<\lambda_{\rm alg}$, the PDA fixed point distributions of the fields $\sH$ and $\hat\sH$ are supported on finite values, predicting a \textit{partial} recovery of the hidden matching with belief propagation, i.e., $0<\mathbb E[\varrho]<1$, see Fig.~\ref{fig:disth}. 

\begin{figure}
    \centering
    \includegraphics[width=\columnwidth]{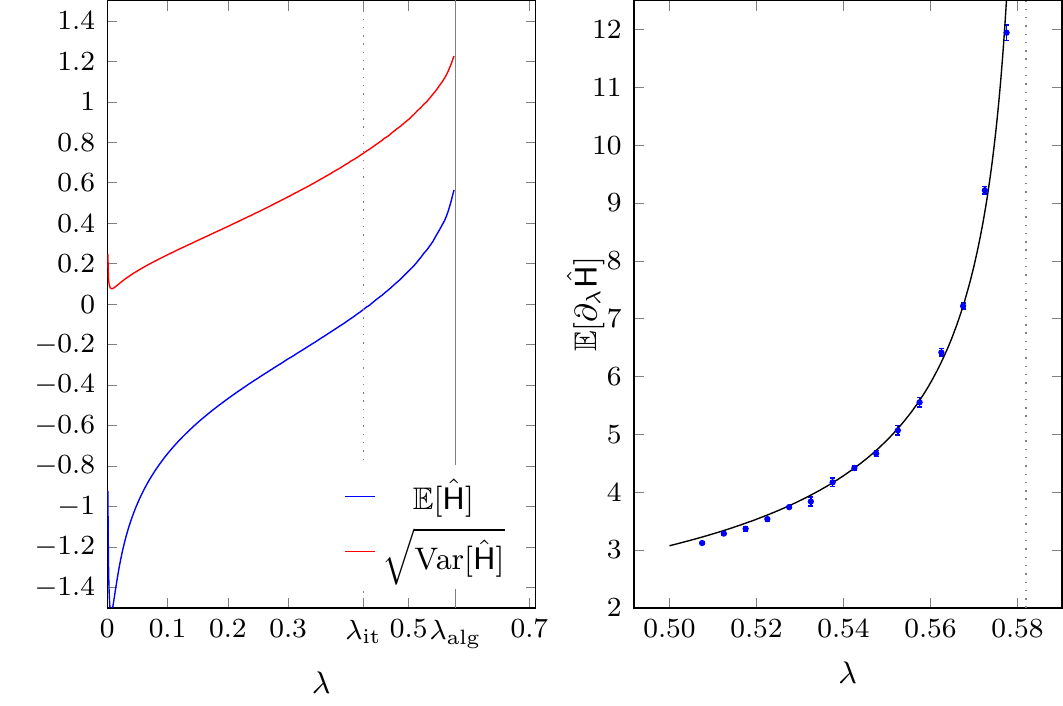}
    \caption{\textit{Left.} Value of $\mathbb E[\hat\sH]$ as function of $\lambda$. The dotted vertical line delimits the thermodynamically stable region of the partial recovery phase, whereas the continuous line corresponds to the algorithmic recovery transition point. \textit{Right.} Numerical derivative $\mathbb E[\partial_\lambda\hat\sH]$ as function of $\lambda$. The smooth line is a fit via a functions $\varphi(\lambda)=a(b-\lambda)^{-\sfrac{1}{2}}$, with best fit values $a=0.882(1)$ and $b=0.582(1)$, slightly larger than the larger value of $\lambda_{\rm alg}=0.578(1)$ estimated via a PDA.}
    \label{fig:mediaH}
\end{figure}

The presence of a first-order transition for $k>2$ can be further corroborated by computing the Bethe free energy, shown in Fig.~\ref{fig:rhoexp1}: the non-trivial fixed point obtained by the PDA for $\lambda<\lambda_{\rm alg}$ has Bethe free energy larger than $f^\star$, free energy corresponding to the planted solution, for $\lambda>\lambda_{\rm it}=0.43(1)$, meaning that such fixed point is thermodynamically unstable in the range $\lambda_{\rm it}<\lambda<\lambda_{\rm alg}$, where therefore $\bfm^s=\bfm^\star$ yet the solution is inaccessible to BPA, which outputs the partial recovery fixed point. The region $\lambda_{\rm it}<\lambda<\lambda_{\rm alg}$ thus marks a \textit{hard phase} where perfect recovery is information-theoretically possible, but belief propagation algorithm does not achieve it. It is conjectured that a much broader class of polynomial algorithms will fail in this region, as escaping the partial recovery fixed point would require an exponentially long time in the size of the problem. Note that similar computational gaps appear, e.g., in the planted XOR-SAT problem \cite{zdeborova2016}, the planted $q$-coloring problem \cite{PhysRevLett.102.238701}, and, more generally, inference problems involving the interaction of more than two variables \cite{krzakala2007}. In these problems, however, the transition typically occurs between a partial recovery (ferromagnetic) phase and a no recovery (paramagnetic) phase.
Finally, the numerical computation of $\mathbb E[\partial_\lambda\hat\sH]$ shows a sharp increase (compatible with a power-law divergence) as $\lambda_{\rm alg}$ is approached, see Fig~\ref{fig:mediaH}, quantitatively expressing the fact that the partial-recovery fixed point becomes unstable at the transition $\lambda_{\rm alg}$.
 
All PDA predictions have been confirmed by numerical simulations performed running a BPA at $\beta=1$ on several instances extracted from the ensemble $\mathcal H_{3,c}^N[\mathrm{Exp}(\lambda),\mathrm{Unif}([0,c])]$ for various values of $N$ and $c=50$. The BPA exhibits a fast convergence, requiring usually less than $10kN$ updates of the fields set, except, as expected, for a slowing down for values of $\lambda$ close to the transition point $\lambda_{\rm alg}$, see Fig.~\ref{fig:rhoexp1}.

\paragraph{The bMAP estimator}
The study of the bMAP can be carried on in a similar manner, relying on the simpler RDEs in Eqs.~\eqref{RDE0}. Just like for the sMAP, it is known that the bMAP exhibits two regimes for $k=2$, namely a {partial recovery phase}, in which $\mathbb E[\varrho]>0$, and a {full recovery phase}, in which $\mathbb E[\varrho]=0$. Remarkably, the relative simplicity of the equations for $k=2$ allowed, in Ref.~\cite{SSZ_2020}, to show that the boundary between the two phases is determined by the condition
\begin{equation}
\mathcal B[p,\hat p]\coloneqq\int\sqrt{p(w)\hat p(w)}\dd w=\frac{1}{\sqrt{c}},\label{criterio}
\end{equation}
where $\mathcal B[p,\hat p]$ is the so-called Bhattacharyya coefficient between the distributions $p$ and $\hat p$ \cite{bhattacharyya1946}. The criterion has been first derived by means of heuristic arguments, and later proved rigorously \cite{ding2021}. Assuming $\hat p=\mathrm{Exp}(\lambda)$ and $p=\mathrm{Unif}([0,c])$, it can be proven in particular that for $c\to+\infty$ an \textit{infinite-order transition} takes place at $\lambda_{\rm alg}=4$, i.e., $\mathbb E[\varrho]$ approaches zero as $\lambda\to4^-$ with all its derivatives \cite{SSZ_2020,ding2021}. Numerical evidences suggest that the transition is continuous for finite values of $c$ as well \cite{SSZ_2020}.

\begin{figure}
    \centering
   \includegraphics[width=\columnwidth]{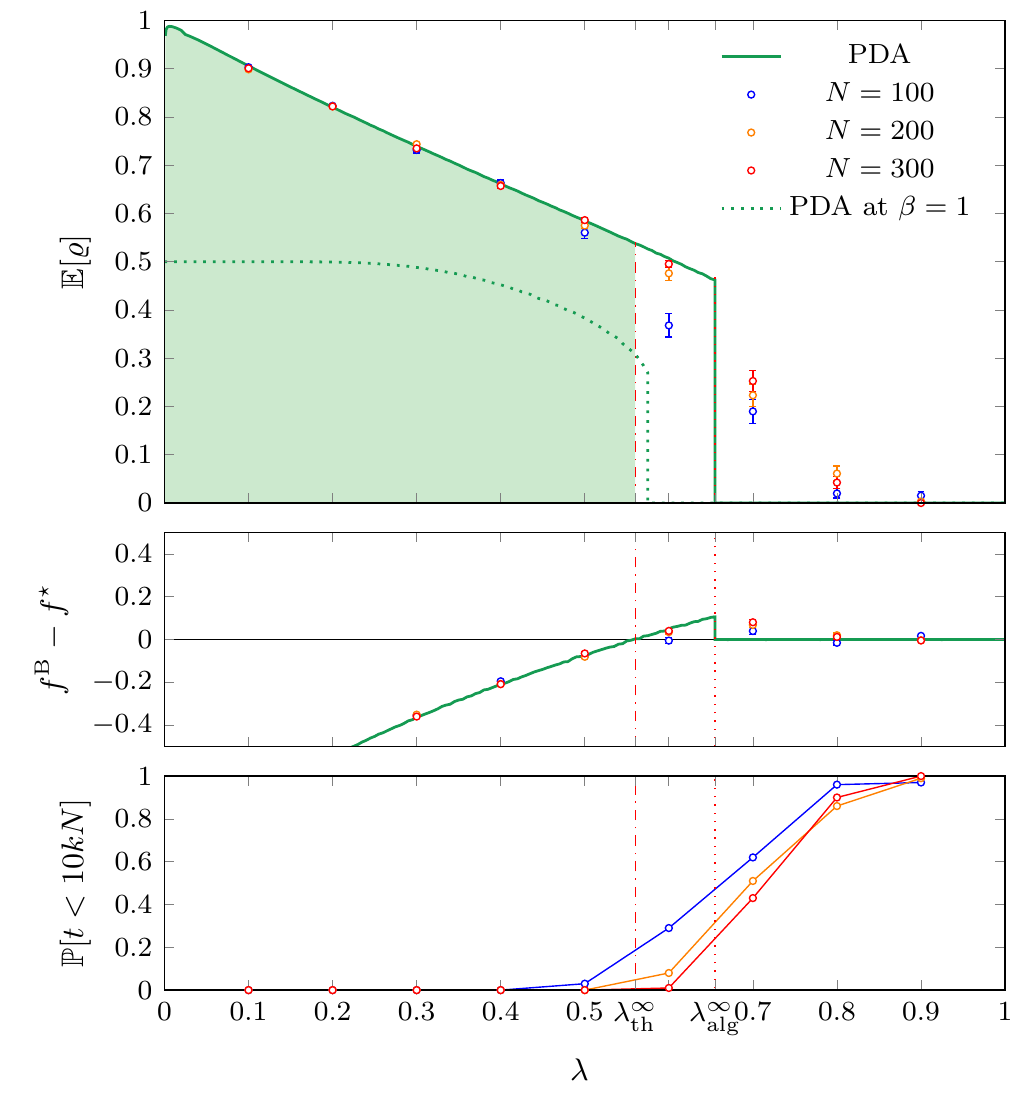}
    \caption{Numerical results for the planted MIMP at $\beta\to+\infty$. Smooth curves are obtained from a PDA solving the RDEs in Eq.~\eqref{RDE0} with $k=3$. We assume that $\hat p=\mathrm{Exp}(\lambda)$ and $p=\mathrm{Unif}([0,c])$. The PDA used a population of $10^5$ fields updated $200$ times for each value $\lambda$ with $c=50$, a value large enough to see no dependence on $c$ of the obtained curves (except for small values of $\lambda$, where we used $c=200$ to avoid finite-$c$ effects near the origin). Dots are obtained by running a BPA on $10^2$ instances of the ensemble $\mathcal H_{3,50}^N[\hat p,p]$. 
    \textit{Top.} Average error $\mathbb E[\varrho]$ for the bMAP: the cavity prediction is compared with the results of numerical simulations. \textit{Center.} Difference between the Bethe free energy provided by the cavity method and the free energy of the planted solution. The fixed point obtained by the PDA is thermodynamically unstable in an interval $\lambda_{\rm alg}^\infty<\lambda<\lambda_{\rm alg}$. \textit{Bottom.} Probability $\mathbb P[t<10kN]$ that the algorithm requires a number of sweep smaller than $10kN$ to reach convergence: it is observed that convergence is never achieved within this number of sweeps for $\lambda<\lambda_{\rm alg}^\infty$.
    }
    \label{fig:rhoexp}
\end{figure}

\begin{figure}
    \centering
   \includegraphics[width=\columnwidth]{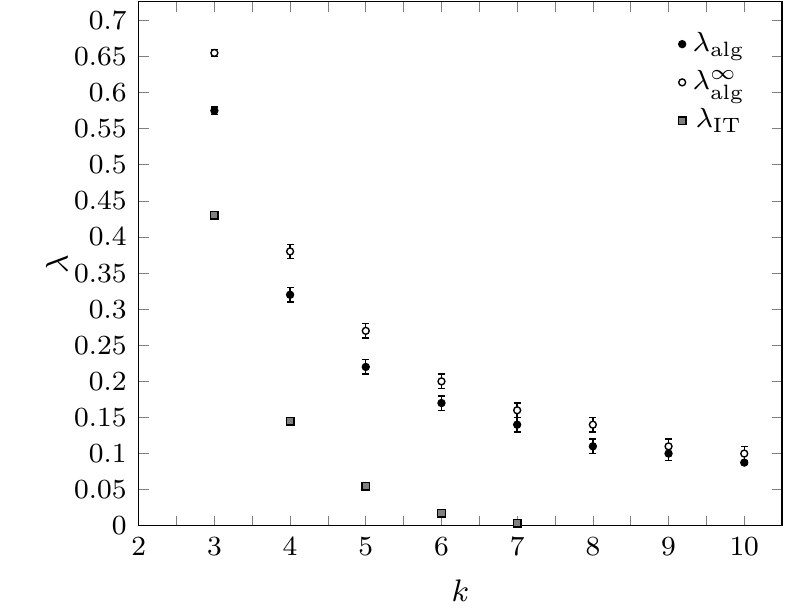}
    \caption{Algorithmic transition point from partial to full recovery phase in the planted MIMP by estimating the sMAP and the bMAP. Results are obtained via a PDA, updating $300$ times a population of $10^5$ fields for each coordination $k$ of hyperedges from $k=3$ to $k=10$. For our numerical simulations, we assumed $c=10^2$, $\hat p=\mathrm{Exp}(\lambda)$ and $p=\mathrm{Unif}([0,c])$. It is observed that larger values of $k$ corresponds to an easier recovery and in particular the partial recovery phase shrinks as $k\to+\infty$. For comparison, we also plot $\lambda_{\rm IT}$ for $k\leq 7$. We observe the region in which it is information-theoretically impossible to fully reconstruct the signal rapidly shrinks to zero as $k$ increases, and we estimate $\lambda_{\rm IT}<0.05$ for $k\geq 8$.
    }
    \label{fig:bigk}
\end{figure}

Let us now consider the problem of estimating the bMAP on a graph obtained from the ensemble $\mathcal H^N_{3,c}[\hat p,p]$, assuming as before $\hat p=\mathrm{Exp}(\lambda)$ and $p=\mathrm{Unif}([0,c])$, and taking the $c\to+\infty$ limit for simplicity. In Fig.~\ref{fig:rhoexp} it is shown that a nontrivial distributional fixed point is obtained for $\lambda<\lambda_{\rm alg}^\infty=0.66(1)$, corresponding to a partial recovery regime, whilst for $\lambda>\lambda_{\rm alg}^\infty$ optimal performances are achieved and $\mathbb E[\varrho]\equiv 0$. Unlike the $k=2$ case, but as observed for the sMAP, the transition is found to be of first order, with a sharp jump in $\mathbb E[\varrho]$ to zero, corroborated by an overshoot of the Bethe free energy with respect to the planted value $f^\star$ in an interval $\lambda_{\rm th}^\infty<\lambda<\lambda_{\rm alg}$, with $\lambda_{\rm th}^\infty= 0.56(1)$. As expected, the performances in terms of the error $\rho$ obtained running the algorithm at $\beta\to+\infty$ are worse than the corresponding at $\beta=1$. In Fig.~\ref{fig:bigk}, we plot the transition points $\lambda_{\rm alg}^\infty$ and $\lambda_{\rm alg}$ estimated by a PDA for values of the coordination of hyperedges $k$ from $3$ to $10$. The results suggest that the difference in $\lambda_{\rm alg}^\infty-\lambda_{\rm alg}$ reduces as $k$ increases.

\begin{figure}
    \centering
   \includegraphics[width=\columnwidth]{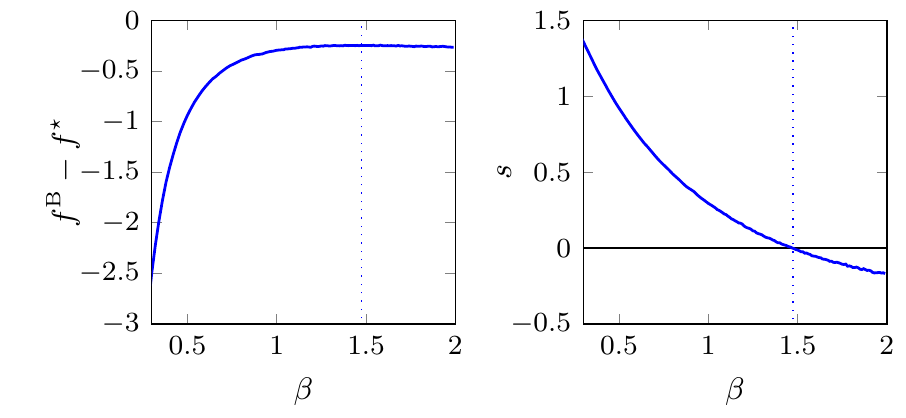}
    \caption{Replica-symmetric estimation of the free energy and of the entropy of the problem as a function of $\beta$ for $\lambda=0.3$ (e.g., inside the partial-recovery region) via a PDA, obtained using $c=50$. Observe that for $\beta>\beta_{\rm dAT}>1$ the entropy takes negative values, a fact that suggests the presence of replica symmetry breaking.
    }
    \label{fig:entropy}
\end{figure}

We have numerically tested the PDA predictions running the BPA on several instances of $\mathcal H^{N}_{3,c}[\mathrm{Exp}(\lambda),\mathrm{Unif}([0,c])]$ for various values of $N$ and assuming $c=50$. Interestingly, the BPA typically \textit{did not} converge within our simulation times for $\lambda<\lambda_{\rm alg}^\infty$: in Fig.~\ref{fig:rhoexp} we plot $\mathbb P[t<10kN]$, probability that the BPA requires a number $t$ of updates of all cavity fields smaller than $10kN$, observing that such probability is estimated to be zero in the partial recovery phase, and decreases to zero in the full-recovery phase. For $\lambda<\lambda_{\rm alg}^\infty$ we stopped the algorithm anyway after $10Nk$ iterations, and computed the error $\varrho$ using the edge set $\hat m_e=\theta\big(\sum_{v\in\partial e}h_{v\to e}\geq\omega_e\big)$, $e\in\mathcal E$: remarkably, this estimator exhibits an overlap with the ground truth which is fully compatible with the value $\mathbb E[\varrho]$ predicted by the PDA, although $\hat\bfm=(\hat m_e)_{e\in\mathcal E}$ is not a matching map as the bMAP should be. The lack of convergence of the algorithm suggests the possibility that the $\beta\to+\infty$ regime within the partial recovery interval lays in a RSB phase. If this is the case, our approach (that assumes the existence of at most one distributional fixed point with finite support) is incorrect. Possibly the simplest consistency test in this direction goes through the computation of the entropy $s(\beta)=\beta^2\partial_\beta f^{\rm B}(\beta)$ as function of $\beta$ \cite{MMR_2005}, a quantity which can be estimated once again using the PDA. Our results are given in Fig.~\ref{fig:entropy}, where both the Bethe free energy $f^{\rm B}(\beta)$ and the entropy $s(\beta)$ are plotted as a function of $\beta$ for a value $\lambda$ in the partial recovery regime: we found that there exists a value $\beta_{\rm dAT}(\lambda)>1$ where the entropy becomes negative, and therefore the replica-symmetric scenario breaks down. By consequence, a proper study of the BP algorithm at $\beta\to+\infty$ would require a replica-symmetry-broken formalism within the partial recovery phase.

\section{The mixed case:\\ the planted mixed MIMP}\label{sec:mixed}
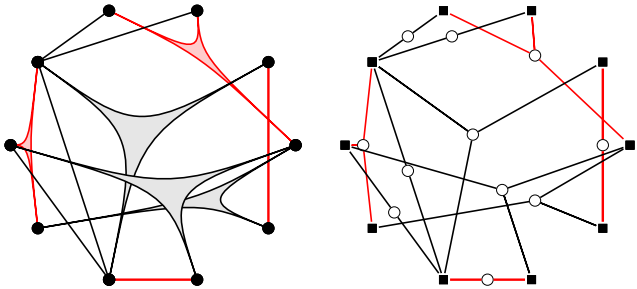
\begin{figure}
    \begin{center}
{ \newcommand{\hyperedgeb}[7][]
{
\draw[#1]($(#2)$) to [out=#5,in=#6] ($(#3)$) to [out=#6, in=#7] ($(#4)$) to [out=#7, in=#5] ($(#2)$);
\node [circle,draw,fill=black,inner sep=2] () at ($(#2)$) {};
\node [circle,draw,fill=black,inner sep=2] () at ($(#3)$) {};
\node [circle,draw,fill=black,inner sep=2] () at ($(#4)$) {};
}

\newcommand{\hyperedge}[4][]
{
\node (C) at (barycentric cs:#2=1,#3=1,#4=1) {};
\draw[#1]($(#2)$) .. controls ($(C)$) and ($(C)$) .. ($(#3)$) .. controls ($(C)$) and ($(C)$) .. ($(#4)$) .. controls ($(C)$) and ($(C)$) .. ($(#2)$);
\node [circle,draw,fill=black,inner sep=2] () at ($(#2)$) {};
\node [circle,draw,fill=black,inner sep=2] () at ($(#3)$) {};
\node [circle,draw,fill=black,inner sep=2] () at ($(#4)$) {};
}

\newcommand{\edge}[3][]
{
\node (C) at (barycentric cs:#2=1,#3=1) {};
\draw[#1]($(#2)$) -- ($(#3)$);
\node [circle,draw,fill=black,inner sep=2] () at ($(#2)$) {};
\node [circle,draw,fill=black,inner sep=2] () at ($(#3)$) {};
}

\newcommand{\hyperedgeF}[4][]
{
\node[circle,draw,fill=white,inner sep=2] (C) at (barycentric cs:#2=1,#3=1,#4=1) {};
\draw[#1,fill=none](#2) -- (C) -- (#3) -- (C) -- (#4);
\node [rectangle,draw,fill=black,inner sep=2] () at ($(#2)$) {};
\node [rectangle,draw,fill=black,inner sep=2] () at ($(#3)$) {};
\node [rectangle,draw,fill=black,inner sep=2] () at ($(#4)$) {};
}

\newcommand{\edgeF}[3][]
{
\node[circle,draw,fill=white,inner sep=2] (C) at (barycentric cs:#2=1,#3=1) {};
\draw[#1,fill=none](#2) -- (C) -- (#3);
\node [rectangle,draw,fill=black,inner sep=2] () at ($(#2)$) {};
\node [rectangle,draw,fill=black,inner sep=2] () at ($(#3)$) {};
}
\tikzset{
    position/.style args={#1:#2 from #3}{
        at=(#3.#1), anchor=#1+180, shift=(#1:#2)
    }
}
\scalebox{0.75}{
 \begin{tikzpicture}[scale=1]
\foreach \x [count=\p] in {0,...,9} {
    \node[draw=none] (v\p) at (-\x*36:2.5) {};};
\edge[very thick,red]{v10}{v2}
\edge[very thick,red]{v3}{v4}
\hyperedge[thick,red,fill=red!20]{v5}{v6}{v7}
\hyperedge[thick,red,fill=red!20]{v8}{v9}{v1}
\hyperedge[thick,black,fill=gray!20]{v1}{v2}{v5}
\hyperedge[thick,black,fill=gray!20]{v4}{v7}{v10}
\hyperedge[thick,black,fill=gray!20]{v6}{v3}{v1}

\edge[thick,black,fill=gray!20]{v4}{v7}
\edge[thick,black,fill=gray!20]{v4}{v6}
\edge[thick,black,fill=gray!20]{v7}{v9}
\edge[thick,black,fill=gray!20]{v7}{v8}
\end{tikzpicture}\qquad
 \begin{tikzpicture}[scale=1]
\foreach \x [count=\p] in {0,...,9} {
    \node[draw=none] (v\p) at (-\x*36:2.5) {};};
\edgeF[very thick,red]{v10}{v2}
\edgeF[very thick,red]{v3}{v4}
\hyperedgeF[thick,red,fill=red!20]{v5}{v6}{v7}
\hyperedgeF[thick,red,fill=red!20]{v8}{v9}{v1}
\hyperedgeF[thick,black,fill=gray!20]{v1}{v2}{v5}
\hyperedgeF[thick,black,fill=gray!20]{v4}{v7}{v10}
\hyperedgeF[thick,black,fill=gray!20]{v6}{v3}{v1}

\edgeF[thick,black,fill=gray!20]{v4}{v7}
\edgeF[thick,black,fill=gray!20]{v4}{v6}
\edgeF[thick,black,fill=gray!20]{v7}{v9}
\edgeF[thick,black,fill=gray!20]{v7}{v8}
\end{tikzpicture}
}}
    \end{center}
    \caption{\textit{On the left}, pictorial representation of a $(2+3)$-hypergraph with a matching (in red) on it. \textit{On the right}, corresponding factor graph: we used the same graphical convention as in Fig.~\ref{fig:hypergraph}.}
    \label{fig:hypergraph2}
\end{figure}

The different nature of the transition in the $k=2$ case and in the $k>2$ case motivated us to consider an ensemble of graphs presenting a mixture of edges and hyperedges, see e.g.~Fig.~\ref{fig:hypergraph2}. We introduce therefore a new ensemble of hypergraphs $\hat{\mathcal H}_{r,c}^N[\hat p,p]$ interpolating between the ensemble $\mathcal H^N_{2,c}[\hat p,p]$ and $\mathcal H_{3,c}^N[\hat p,p]$, depending on two absolutely continuous distributions $\hat p$ and $p$, an integer $N\in\mathds N$, a real number $c\in\R^+$ and on a parameter $r\in[0,1]$, $rN\in\mathds N$. In this ensemble, a graph with $6N$ vertices is constructed as follows.
\begin{enumerate}
\item The vertex set $\mathcal V$ is divided into two subsets, namely $\mathcal V_2$, containing $6(1-r)N$ vertices, and $\mathcal V_3$, containing $6rN$ vertices. Vertices within $\mathcal V_2$ are linked in pairs, uniformly choosing a matching amongst all possible perfect pairing in the set. Vertices within $\mathcal V_3$ are grouped in $3$-plets, each joined by a hyperedge, uniformly choosing a partition in triplets amongst all possible ones. The resulting edge set $\mathcal M_0$, will play the role of \textit{planted} matching and is therefore a mixture of $3(1-r)N$ edges and $2rN$ hyperedges. Each planted edge or hyperedge $e\in\mathcal M_0$ is associated to a weight $w_e$, extracted with probability density $\hat p$ independently from all the others.
\item Given all possible $\binom{6N}{3}-2rN$ $3$-hyperedges not in $\mathcal M_0$, we add each of them with probability $2cr(6N)^{-2}$. Similarly, we add each of the $\binom{6N}{2}-3(1-r)N$ possible edges not in $\mathcal M_0$ with probability $c(1-r)(6N)^{-1}$. We denote $\mathcal E_0^{\rm np}$ the set of newly added edges or hyperedges, we call them \textit{non-planted}. For large $N$, each vertex in the constructed graph has an outgoing planted edge (planted hyperedge, respectively) with probability $1-r$ (with probability $r$, respectively); in addition to this, it has, on average, $cr$ outgoing non-planted hyperedges and $(1-r)c$ outgoing non-planted edges, so that the obtained graph has overall on average $3c(1-r)N$ non planted edges and $2crN$ non planted $3$-hyperedges. Each non-planted edge or hyperedge $e\in\mathcal E_0^{\rm np}$ is associated to a weight $w_e$, extracted with probability density $p$, independently from all the others.
\end{enumerate}
The rules given above are such that, for $r=0$ we sample an element of the ensemble $\mathcal H_{2,c}^{6N}[\hat p,p]$, whilst $r=1$ corresponds to a graph of $\mathcal H_{3,c}^{6N}[\hat p,p]$. The analysis in Section~\ref{sec:ens} and Section~\ref{sec:bp} can be repeated for the newly introduced ensemble and, in particular, we can implement a BPA in the same form as in Eqs.~\eqref{bp2} on a factor graph in which variable nodes have coordination $2$ if  corresponding to edges, and coordination $3$ if corresponding to hyperedges, see Fig.~\ref{fig:hypergraph2}. For the sake of brevity, we do not repeat the derivation here. We denote as $\sK$ and $\sK'$ two random variables with distribution 
\begin{subequations}
\begin{align}
\mathbb P[\sK=k]&=(1-r)\delta_{k,2}+r\delta_{k,3}\\
\mathbb P[\sK'=k]&\comprimi=\frac{k\mathbb P[\sK=k]}{\mathbb E[\sK]}.
\end{align}
\end{subequations}
Defining $\gamma_\sK\coloneqq c\mu\hat\mu^{\sK-1}$, the effect of the pruning can be condensed in the quantities
\begin{subequations}\begin{align}
\hat q&=1-\e^{-\mathbb E[\gamma_\sK q^{\sK-1}]},\\
q&=\mathbb E[\hat q^{\sK'-1}],
\end{align}
\end{subequations}
where $q$ and $\hat q$ have the same meaning as corresponding quantities in Section~\ref{sec:bp}. The average reconstruction achieved by the BP algorithm on a graph of this ensemble can be written then in terms of random variables $\sH$ and $\hat\sH$ satisfying RDEs formally identical to the ones in Eqs.~\eqref{RDE1} once $k$ is replaced by the random variable $\sK'$ and $\sZ\sim\mathrm{ZTPoiss}(\mathbb E[\gamma_\sK q^{\sK-1}])$. In particular, the average error is
\begin{equation}\comprimi
    \mathbb{E} [\varrho] = \frac{\hat\mu}{2} \mathbb{E} \left[\hat q^{\sK} \theta\left(\hat\sOmega-\sum_{v=1}^\sK \hat{\sH}_v  \right)\right] + \frac{\hat\mu}{2} \mathbb{E} \left[\gamma_\sK q^{\sK} \theta\left(\sum_{v=1}^
    \sK\sH_v - \sOmega\right) \right].
\end{equation}

\begin{figure}
    \centering
    \includegraphics[height=0.8\columnwidth]{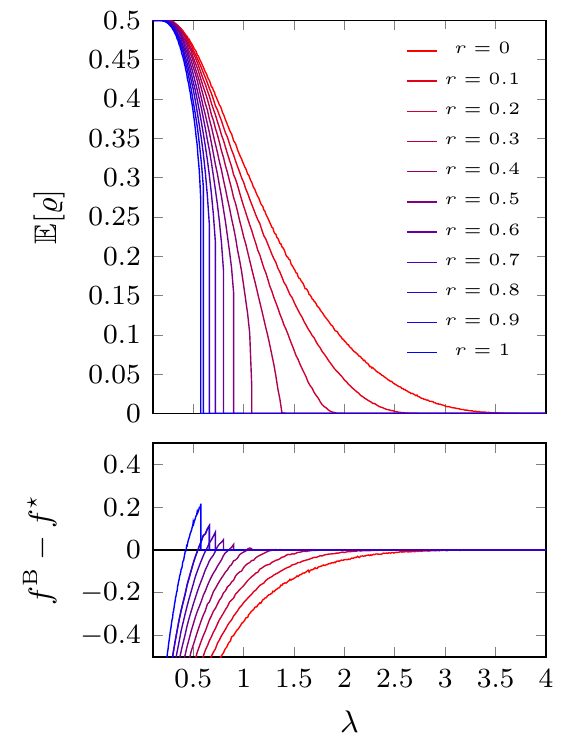}
    \includegraphics[height=0.8\columnwidth]{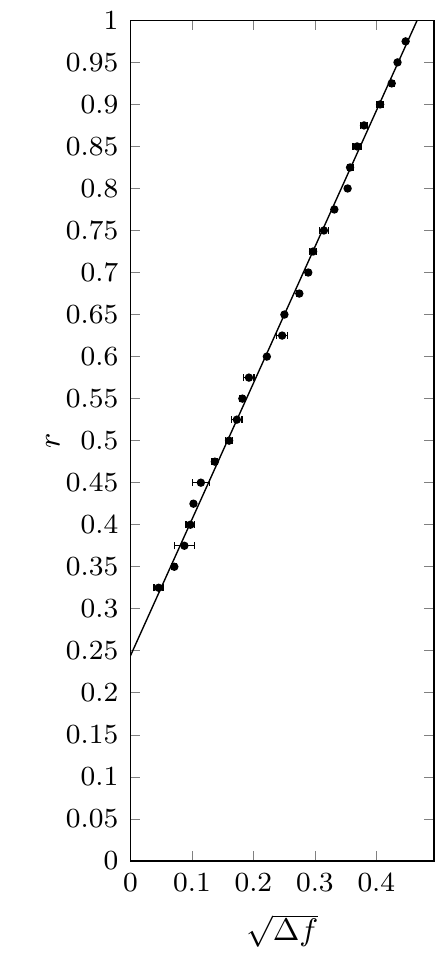}
    \caption{Numerical results for the mixed planted MIMP at $\beta=1$. The curves are obtained via a PDA at $\beta=1$. We assume that $\hat p=\mathrm{Exp}(\lambda)$ and $p=\mathrm{Unif}([0,c])$. The PDA used a population of $10^5$ fields updated $300$ times for each value $\lambda$ with $c=100$. \textit{Left top.} Average error $\mathbb E[\varrho]$ for various values of $r$ at $\beta=1$. \textit{Left bottom.} Difference between the Bethe free energy obtained via the PDA and the free energy of the planted solution. \textit{Right.} Relative amount $r$ of edges and hyperedges as function of the square root of the maximal overshoot of the Bethe free energy. A null overshoot is estimated via linear fit on values of $\sqrt{\Delta f}$ (continuous line) at $r_0=0.244(4)$. For $r<0.3$ the numerical value of $\max_\lambda f^{\rm B}$ was indistinguishable from numerical fluctuations around $f^\star=1$ and we omitted therefore the corresponding data points.}
    \label{fig:errmix}
\end{figure}

As in the pure case, we numerically solved the RDEs for the mixed case and we considered $\hat p=\mathrm{Exp}(\lambda)$ and $p=\mathrm{Unif}([0,c])$ in the limit $c\to+\infty$. The value of the average error $\mathbb E[\varrho]$ for $\beta=1$ is given in Fig.~\ref{fig:phaseB1}, that visually renders the crossover between a first order transition at $r=1$ and a continuous transition at $r=0$. This is more clearly visible in Fig.~\ref{fig:errmix}, where the value of $\mathbb E[\varrho]$ is plotted as function of $\lambda$ for different values of $r$. In Fig.~\ref{fig:errmix} we plot the overshoot $\Delta f\coloneqq \max_\lambda f^{\rm B}-f^\star$ as a function of $r$: we numerically find $\Delta f=a(r-r_0)^2\theta(r-r_0)$, with $r_0=0.244(4)$. We therefore conjecture  that the transition becomes of second order at $r=r_0=0.244(4)$. 

\section{Conclusions}\label{sec:conclusions}
We have studied the problem of inferring a (weighted) planted MIMP hidden in a random $k$-hypergraph, relying on the information provided by the topology and the weights on the edges. In particular, the weights of the hidden structure were assumed to be randomly distributed according to an absolutely continuous density $\hat p$, whereas all the remaining weights follow a different absolutely continuous density $p$. Under the assumption of locally tree-like structure of the graph and fast-decaying correlations, we wrote down a message-passing algorithm to estimate the marginal probabilities of each edge of belonging to the hidden matching. The performance of the algorithm was studied by numerically solving a set of recursive distributional equations via a population dynamics algorithm. We have focused in particular on two different estimators for the hidden matching constructed from the obtained marginals, namely the sMAP (which is Bayes optimal with respect to the Hamming distance with the hidden matching) and the bMAP (corresponding to the perfect matching with highest overall likelihood). For both estimators, and in the large-graph-size limit, a phase transition takes place with respect to the signal intensity between a phase in which full recovery of the hidden structure is feasible and a phase in which instead only partial recovery is accessible. Remarkably, the transition is found to be of \textit{first order} for $k>2$, in contrast with the $k=2$ case where the transition is continuous, implying that there is a regime of the signal-to-noise ratio where the full recovery of the signal is hard and a computational gap appears. Moreover, in the case of belief propagation for the bMAP, the partial-recovery phase is characterised by lack of convergence of the algorithm, which is typically unable to output a perfect matching, although an early stopping provides a set of edges correlated with the hidden signal whose size is correctly predicted by the RDEs: we have shown that this algorithmic hardness is likely due to the presence of an RSB phase in the phase diagram. 

Although the main properties of the problem can be investigated via a PDA, an explicit instability criterion for determining the transition point $\lambda_{\rm alg}$ at $k>2$ is still missing and left for future investigations.

Finally, we have analysed a mixed model in which both edges and $3$-hyperedges coexist. We have shown that the aforementioned phase transition persists in the mixed settings, and interpolates between the continuous transition for the pure $k=2$ case and the first-order transition (with computational gap) of the $k=3$ case. We have presented numerical evidences, in particular, that the presence of a finite fraction of edges in the hypergraph makes the transition of second order. This phenomenology is reminiscent of what is observed in other planted problems, in particular the spiked mixed matrix-tensor model \cite{Sarao2020}, in which a mixture of two-body and $p$-body interaction terms allows to interpolate between a second order transition and a first order transition: note however that in such problems the transition occurs between a no recovery phase and a partial recovery phase.

\subsection*{Acknowledgments}
The authors are grateful to Guilhem Semerjian, Stefano Sarao Mannelli and Pierfrancesco Urbani for useful discussions.

\appendix
\section{Expression for the error in the bMAP for the planted $k$-MIMP}\label{app1}
In this Appendix we prove Eq.~\eqref{eq:rho0} by showing that at $\beta\to+\infty$,
\begin{equation}\textstyle
\hat q^k\mathbb P\left[\hat\sOmega\geq \sum_{u=1}^k\hat\sH_u\right]=q^k\gamma\mathbb P\left[\sOmega\leq \sum_{u=1}^k\sH_u\right]\label{app:eq:A}
\end{equation}
by straightforwardly generalising the arguments in Ref.~\cite{SSZ_2020} for the $k=2$ case. Eq.~\eqref{RDE0b} implies
\begin{equation}
\mathbb P\left[\hat\sOmega-\sum_{u=1}^{k-1}\hat\sH_u\geq x\right]=\frac{\mathbb P[\sH\geq x]}{1-\hat q+\hat q\mathbb P[\hat\sH\geq x]}
\end{equation}
and therefore, in the partial recovery phase,
\begin{multline}\textstyle
\mathbb P\left[ \hat\sOmega-\sum_{u=1}^{k}\hat\sH_u\geq 0\right]=\\\textstyle
=-\int_{-\infty}^{+\infty}\mathbb P\left[\hat\sOmega-\sum_{u=1}^{k-1}\hat\sH_u\geq x\right]\partial_x\mathbb P[\hat\sH\geq x]\dd x\\
=-\frac{1}{\hat q}\int_{-\infty}^{+\infty}\mathbb P\left[\sH\geq x\right]\partial_x\ln\left(1-\hat q+\hat q\mathbb P[\hat\sH\geq x]\right)\dd x\\
=\frac{1}{\hat q}\int_{-\infty}^{+\infty}\partial_x\mathbb P\left[\sH\geq x\right]\ln\left(1-\hat q+\hat q\mathbb P[\hat\sH\geq x]\right)\dd x.
\end{multline}
We write now
\begin{multline}\comprimi
\mathbb P[\hat\sH>x]=\frac{1-\hat q}{\hat q}\sum_{k=1}^\infty \frac{1}{n!} \left(q^{k-1}\gamma\mathbb P\Big[\sOmega-\sum_{u=1}^{k-1}\sH_u\geq x\Big]\right)^n\\
=(1-\hat q)\frac{\exp\left(q^{k-1}\gamma \mathbb P\Big[\sOmega-\sum_{u=1}^{k-1}\sH_u\geq x\Big]\right)-1}{\hat q}\\
=\frac{\exp\left(-q^{k-1}\gamma \mathbb P\Big[\sOmega-\sum_{u=1}^{k-1}\sH_u\leq x\Big]\right)-1+\hat q}{\hat q}
\end{multline}
so that $\ln(1-\hat q+\hat q\mathbb P[\hat\sH\geq x])=-q^{k-1}\gamma \mathbb P[\sOmega-\sum_{u=1}^{k-1}\sH_u\leq x]$. Using the fact that $\hat q^{k-1}=q$, then $\frac{q^{k-1}}{\hat q}=\frac{q^k}{\hat q^k}$ and Eq.~\eqref{app:eq:A} follows. 

\begin{figure}
\vspace{1mm}
    \centering
    \includegraphics[width=\columnwidth]{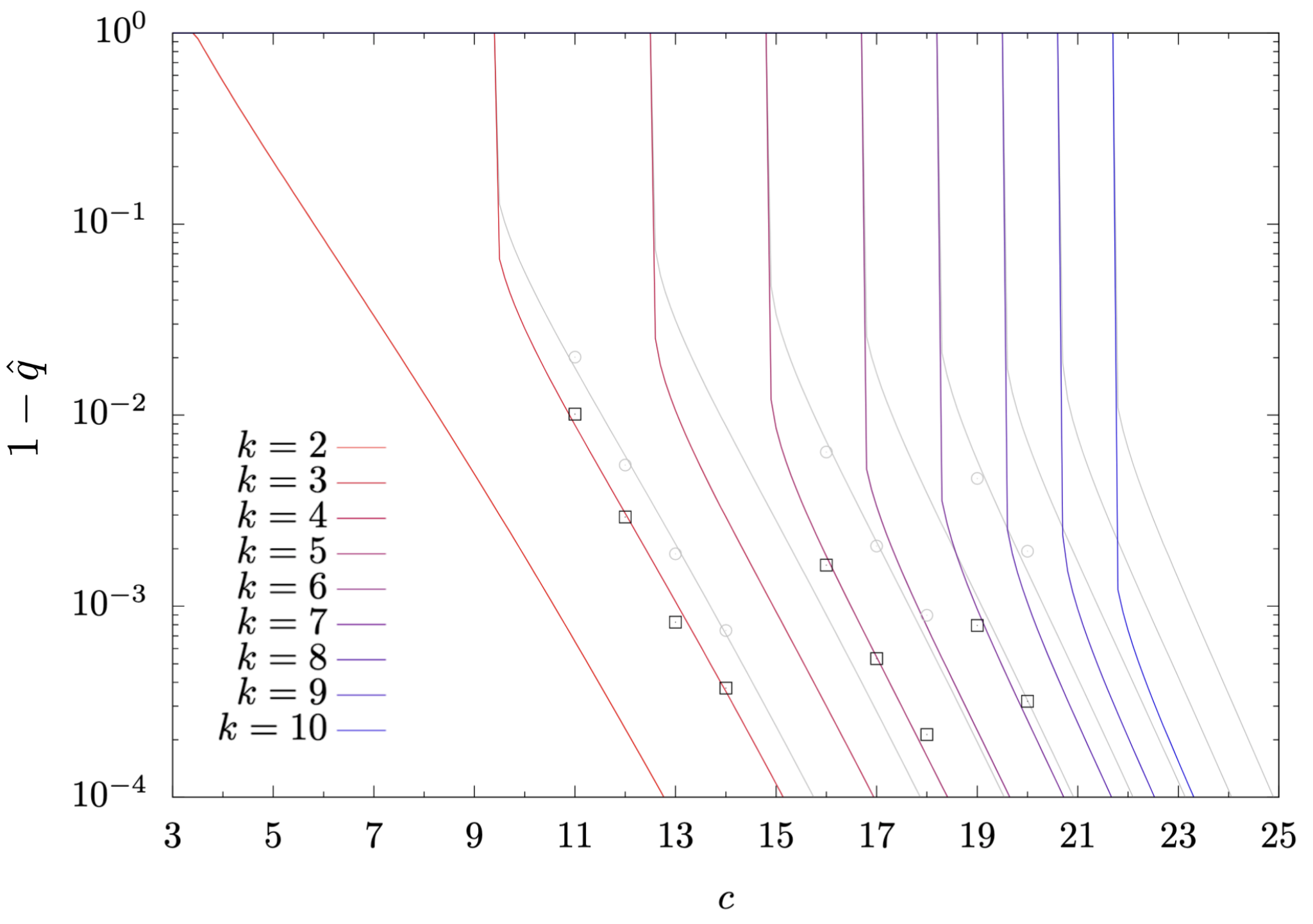}
    \caption{In color, analytic prediction of the probability $1-\hat{q}$ for a planted edge to be pruned  for $\lambda=0.1$. The probability $\hat{q}$ is calculated recursively using Eq.~\eqref{eq:qhat} for different values of $k$. The values for the probability $1-q$ for a non-planted edge to be pruned are in grey. The values $1-\hat{q}$ (resp. $1-q$), obtained by pruning $10^2$ instances of the ensemble $\mathcal H_{k,c}^{2000}[\hat p,p]$ for $k=3,5,7$ and various values of $c$, are shown as black squares (resp. grey dots). The $k=2$ reduction of the formula has been verified with a BPA in \cite{SSZ_2020}.}
    \label{fig:qhat}
\end{figure}

\begin{figure}[t]
    \centering
    \includegraphics[width=\columnwidth]{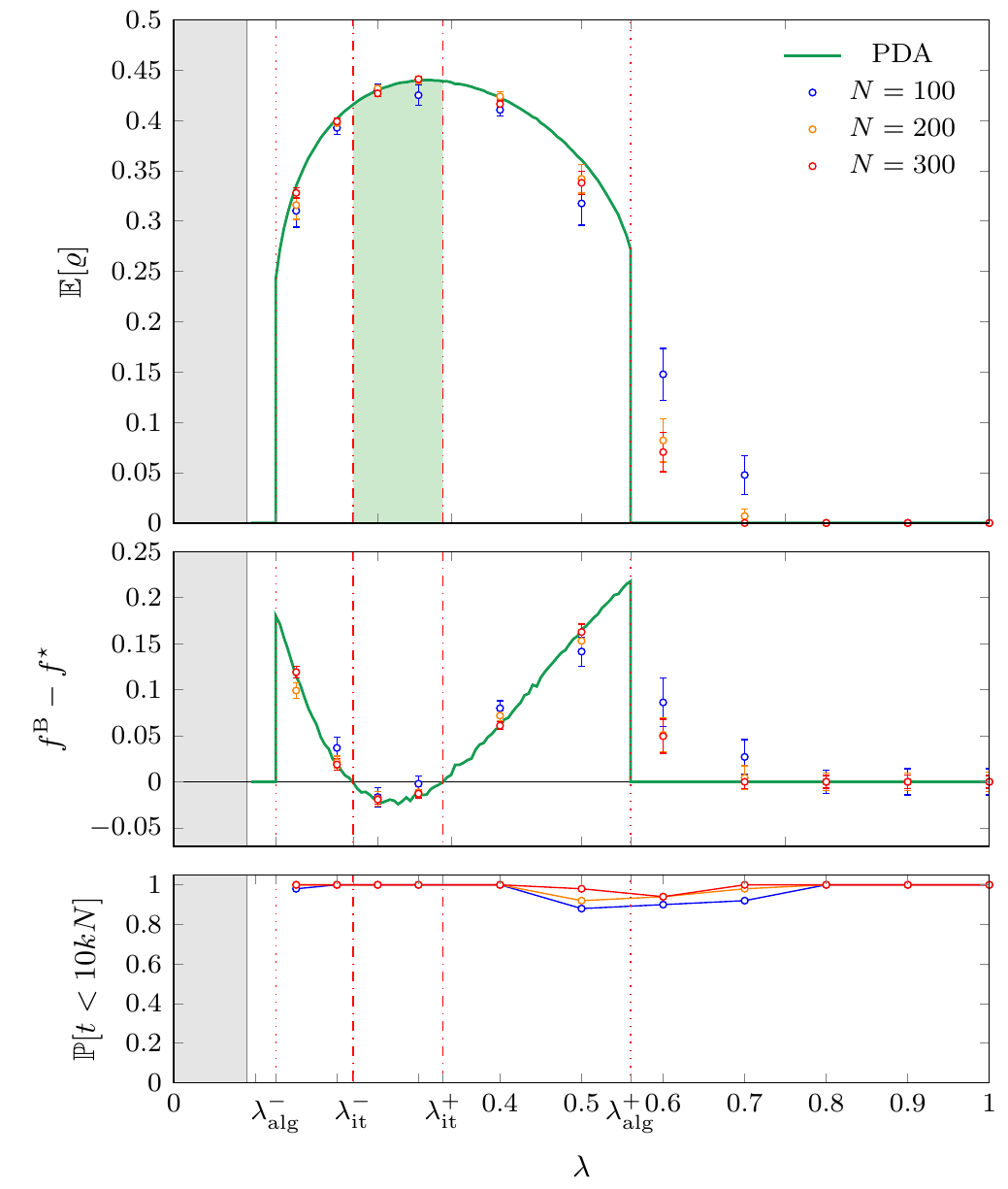}
    \caption{Numerical results on the bMAP for the planted MIMP. Smooth curves are obtained from a PDA solving the RDEs in Eq.~\eqref{RDE1} at $\beta=1$ with $k=3$. We assume that $\hat p=\mathrm{Exp}(\lambda)$ and $p=\mathrm{Unif}([0,c])$. The PDA used a population of $10^5$ fields updated $200$ times for each value $\lambda$ with $c=10$. Dots are obtained by running a BPA on $50$ instances of the ensemble $\mathcal H_{3,10}^N[\hat p,p]$. \textit{Top.} Average error $\mathbb E[\varrho]$ for the sMAP: the PDA prediction is compared with the results of numerical simulations. Note that complete pruning allows for full recovery for $\lambda<\lambda_{\rm pr}=0.08975(5)$ (gray interval). \textit{Center.} Difference between the Bethe free energy obtained via the PDA and the free energy of the planted solution. The fixed point obtained by the PDA is thermodynamically stable in the interval $\lambda_{\rm it}^-<\lambda<\lambda_{\rm it}^+$ (green region), properly contained in the partial recovery region $\lambda_{\rm alg}^-<\lambda<\lambda_{\rm alg}^+$. \textit{Bottom.} Probability $\mathbb P[t<10kN]$ that the algorithm requires a number of sweep smaller than $10kN$ to reach convergence: particularly hard instances appear for $\lambda\simeq \lambda_{\rm alg}^\pm$.}
    \label{fig:c10}
\end{figure}

\section{Recovery in the finite-$\boldsymbol c$ case}\label{app:sparse}

In this Appendix, we present some results on random weighted hypergraphs from the ensemble $\mathcal H_{k,c}^N[\hat p,p]$, described in Section~\ref{sec:ens}, with finite values of the average connectivity parameter $c$.
As anticipated, the overall picture is similar to the one described for $c\to+\infty$, with the additional remark that the sparse nature of the graph can guarantee a partial or full recovery of the signal by simple pruning, as discussed in the main text.
Fig.~\ref{fig:qhat} shows the analytic prediction of the probability that a planted (resp.~non-planted) edge or hyperedge is removed during the pruning procedure, $1-\hat{q}$ (resp.~$1-q$), introduced in Section~\ref{sec:bp}, as a function of the average connectivity parameter $c$. Recall that the relation between the two probabilities is $q=\hat{q}^{k-1}$, so that for $k=2$ we have $q=\hat q$.

Assuming, as in the numerical experiment of the main text, $\hat p=\mathrm{Exp}(\lambda)$ and $p=\mathrm{Unif}([0,c])$, in Fig.~\ref{fig:qhat} we observe that there exists a distinct critical value $c^\star_{k,\lambda}$ such that for $c<c^\star_{k,\lambda}$ topological recovery of the perfect matching occurs. The value $c^\star_{k,\lambda}$ grows as $k$ increases; when a leaf is identified, the hyperedge it belongs to is removed along with the hyperedges incident to its remaining $k-1$ endpoints (the higher $c$ is, the more incident hyperedges are removed for each leaf that is identified). Moreover, for $k>2$ there is a sharp jump at $c^\star_{k,\lambda}$ between topological recovery $q=\hat{q}=0$ and values of $q$ and $\hat q$ close to $1$; this jump is not present for $k=2$ where instead the transition is continuous.

In Fig.~\ref{fig:c10} we present the results of solving the RDEs for $\beta=1$, $k=3$ and $c=10$. Unlike the large $c$ case, we observe not one but two sharp transitions in $\mathbb E[\varrho]$ that are between the partial and full recovery phases which correspond to $\mathbb E[\varrho]=0$ and $0<\mathbb E[\varrho]<1$, respectively. As expected, we can fully recover the planted matching for any $0<\lambda<\lambda_{\rm pr}$, interval where the complete pruning of the graph is possible. Both transitions are analogous to the one observed in the large $c$ case where we see a sharp jump in $\mathbb E[\varrho]$ from partial to full recovery of the planted matching. For $c=10$ the jumps are observed at some values $\lambda_{\rm alg}^\pm$, so that for $\lambda_{\rm alg}^-<\lambda<\lambda_{\rm alg}^+$ the cavity fields are supported on finite values and we have $0<\mathbb E[\varrho]$, i.e., a partial recovery of the hidden matching. Outside the interval, on the other hand, full recovery is achieved. The PDA predictions are confirmed by numerical simulations running a BPA at $\beta=1$ for various graph sizes $N$, averaging over multiple instances from the ensemble $\mathcal H_{3,10}^N[\mathrm{Exp}(\lambda),\mathrm{Unif}([0,10])]$. Moreover, the Bethe free energy exhibits the same phenomenology as for large $c$: the non-trivial fixed point is stable in an interval $(\lambda_{\rm it}^-,\lambda_{\rm it}^+)\subset(\lambda_{\rm alg}^-,\lambda_{\rm alg}^+)$. As for the large $c$ simulations in the main text, BPA converges fast except for values of $\lambda$ close to the transition points $\lambda_{\rm alg}^\pm$.

\bibliography{hypermatching}
\end{document}